\documentstyle[aps,prb,epsfig,psfig,epsf]{revtex}
\begin{document}
\setlength{\parindent}{0em}
\draft
\title{Parisi--Symmetry of the Many--Body Quantum Theory of
randomly interacting fermionic systems}
\author{R. Oppermann$^{1,3}$ and B. Rosenow$^2$}
\address{
$^1$Service de physique th\'eorique, CE Saclay, F--1191 Gif--sur--Yvette,
{\it France}\\  
$^2$ MPI f\"ur Kernphysik, D--69029 Heidelberg, {\it Federal Republic of  
Germany}\\
$^3$ Institut f. Theoret. Physik, Univ. W\"urzburg D--97074
W\"urzburg, {\it Federal Republic of Germany}}
\date{\today}
\maketitle
\begin{abstract}
We show that fermion systems with random interactions lead to strong 
coupling between glassy order and fermionic correlations, which culminates in 
the implementation of Parisi replica permutation symmetry breaking (RPSB) 
in their zero temperature quantum field theories. Precursor effects, 
setting in below fermionic Almeida--Thouless lines, become stronger as the 
temperature decreases and play a crucial role for many physical 
properties within the entire low temperature regime. 
The Parisi ultrametric structure is shown to determine the dynamic 
behaviour of fermionic correlations (Green's functions) for large times and 
for the corresponding low energy excitation spectra, which is predicted to 
affect transport properties in metallic (and superconducting) spin glasses. 
Thus we reveal the existence and the detailed form of a number of 
{\it quantum--dynamical fingerprints of the Parisi scheme}. 
These effects, being strongest as 
$T\rightarrow 0$, are contrasted with the replica--symmetric nature of the 
critical field theory of quantum spin glass transitions at $T=0$, which 
display only small corrections at low T from replica permutation symmetry 
breaking (RPSB). RPSB--effects moreover appear to influence the loci of 
the ground state transitions at $O(T^0)$ and hence the phase diagrams.\\ 
From explicit solutions for arbitrary temperatures we also find a new 
representation of the zero temperature Green's function. This leads to a
map of the fermionic (insulating) spin glass solution to the local limit 
solution of a {\it Hubbard model with a random repulsive interaction}. 
This map exists at any number of replica symmetry breaking steps K. 
We obtain the distribution of the Hubbard interaction fluctuation and its 
dependence on the order of RPSB. A generalized mapping between metallic 
spin glass and random U Hubbard model is conjectured. We also suggest 
that the new representation of the Green's function at $T=0$ can be used for 
generalizations to superconductors with spin glass phases. 
Further generalizations of the fermionic Ising spin glass to models with 
additional spin and charge quantum--dynamics occurring in metallic spin 
glasses, or due to Coulomb effects including crossover 
from 4--state per site to effectively 3--state per site models in the 
limit of infinite repulsive Hubbard coupling are briefly considered. 
We compare our spin glass results with recent $d=\infty$ (clean) Hubbard 
model analyses, paying particular attention to the common role of the 
corresponding Onsager reaction fields. 
\end{abstract}

\pagebreak
\section{Introduction}
In this paper we emphasize and explore the implementation of Parisi replica 
permutation 
symmetry breaking \cite{parisi79,parisi1,parisi2,BiYo,rammal} in the 
quantum theory 
of fermionic systems with random interactions. We present details of the 
derivation of a many body quantum theory with ultrametric structure \cite{epl}
and we discuss the results in context of the tricritical phase diagram, 
which is given in the subsequent paper II.\\
Calculations based on the 4--state per site extension of the standard
SK--model provide the framework for detailed information that helps to
analyze a variety of more complicated fermionic spin-- and charge--glass
quantum models.\\
Spin glasses have been described for a long time by models that isolated    
glassy magnetic phenomena from other physical phenomena. 
The source of the frustrated magnetic interaction often hidden in the
RKKY--interaction of metallic systems, this modelling was meant to apply 
to cases that allow to study glassy magnetic order and electronic 
transport in such a semi--independent way. 
A first microscopic diagrammatic approach to describe at the same footing
the formation of spins and the calculation of their frustrated interaction
was designed by Hertz \cite{Hertz76} for a system of randomly placed strong
Hubbard U centers. A way of phenomenological treatment of spin glass order
and transport theory can be found in the book of Fischer and Hertz 
\cite{FiHe}.
A recent microscopic study of the formation of magnetic moments in models
of Hertz' type was made by Sachdev \cite{subirmm}.\\
Quantum spin glass models possess a genuine richness in their low 
temperature phase diagrams that is due to the entanglement of a many body
interaction and the randomness contained in the couplings. The mixing of
spin and charge fluctuations leads to new experimentally observable phenomena
and to fundamentally new structures of the many body theory.\\
These systems link phase transitions of remarkably different character: 
zero temperature so--called quantum phase transitions displaying quantum 
dynamical effects, but also interesting types of thermal tricritical 
transitions \cite{brro}, which mix spin-- and charge--fluctuations and 
also an interference of spin glass features in charge response and transport 
properties. The symmetry classification of these QPT's is totally 
different from that of thermal phase transitions and also does not appear
to resemble, for example, the $T=0$ QPT universality classes of Anderson 
localization.\\
We elaborate details of the manifestation of Parisi symmetry
in a $T=0$ quantum theory \cite{epl} of the fermionic Ising spin glass.
First extensions to models that possess, beyond the ubiquituous fermion 
quantum dynamics, also spin-- and charge--quantum dynamics are also 
considered. We derive the long--time behaviour of the fermion propagator,
which proves to be one of the ideal quantities that represent 
quantum--dynamically the Parisi replica permutation symmetry breaking in 
theories at $T=0$ and at low temperatures. The order parameter
function $q(x)$ \cite{parisi79,parisi1,parisi2} is static and displays
its nontrivial features only on an interval of order $T$. It needs the
linear susceptibility $\chi=\beta(\tilde{q}-\int_0^1 q(x))$, where 
the spin autocorrelation function $\tilde{q}$ remains static for any 
multi--state Ising spin glass, to incorporate an $O(1)$--effect within a
correlation function. The fermion correlations describe the fermionic
spin glasses in a more detailed way than spin-- and charge--correlations.
The latter can however all be obtained from the former ones. Even for 
spin--static models the fermionic correlations, beginning with the fermion
propagator, yield a detailed quantum--dynamical description of RPSB.
In addition, this is reminiscent of the classical--dynamical theories of 
spin glasses \cite{Zippelius,sompolinsky,horner} with imposed 
Glauber--dynamics and despite the fact that the latter was invented to 
avoid the replica trick thus producing a dynamical image of the classical
spin glass solution, which indeed represented the Parisi solution.
It is this analogy which we emphasize by elaborating in this paper the
{\it quantum--dynamical image} of the Parisi solution for the fermionic
Ising spin glass, which is spin-- and charge--static but has fermion 
dynamics exclusively determined from the Hamiltonian.
Excitation spectra encountered in one-- and two--fermion correlators are
seen to be determined by RPSB in a qualititative and quantitative way.\\
While most of our results are given for the 4--state fermionic Ising spin 
glass. The influence of RPSB on the $T=0$--quantum theory emerges in a 
general and apparently exists in a model--independent way. This clearly 
calls for renewed efforts to find RPSB in the other domains of disordered 
electronic systems such as the theory of Anderson--, 
Anderson--Mott--, and similar localization transitions. This is beyond the 
scope of this paper,
but the role of Parisi symmetry in the (insulating) fermionic Ising spin 
glass together with the Ward identity for charge conservation, which we 
work out here, indicates the possible relevance of this symmetry.

\section{Models of Randomly Interacting Systems} 
The most fundamental and intensively studied model of random interaction is 
the Ising model with fully frustrated spin interaction. As any spin model
it can be described in a grand canonical description with appropriate 
imaginary chemical potential. In this way the standard SK Ising 
spin glass model is viewed as a fermionic spin glass, given on 
special points of the imaginary axis of the complex 
$\mu$--plane. On continuing to real chemical potentials one converts the 
Sherrington Kerkpatrick--model into a physical fermionic Ising spin glass. 
As an interacting disordered fermion model the Hamiltonian appears to be simple.
On one hand it only takes into account a random magnetic interaction, 
which, after partial particle--hole transformation is equivalent to a random 
charge--charge interaction and a magnetic field. It allows however for a 
spin--charge coupling due to redistribution among magnetic and 
nonmagnetic states, which depends strongly on the particle pressure and 
hence on the effective spin density.\\ 
From the point of view of an interacting disordered fermion system 
we demonstrate the way in which the known instability against
replica symmetry breaking leads to a quantum field theory with Parisi 
symmetry. A large part of this paper is devoted to explain this by detailed
results for the infinite--range fermionic Ising spin glass. 
Thus the theory presented here starts out from the  
Hamiltonian, which superficially coincides with the classical
SK--model, but its spins are defined as the fermionic objects
\begin{equation}
\sigma\equiv n_{\uparrow}-n_{\downarrow}, \quad 
n_{\sigma}\equiv a^{\dagger}_{\sigma}a_{\sigma}\quad {\rm with}\quad \{
a_{i \alpha}, a_{j \beta} \} =0, \{ a^{\dagger}_{i \alpha},
a_{j \beta} \} = \delta_{i j} \delta_ {\alpha \beta}.
\end{equation}
This introduces a new class of observables and correlations which involve
odd numbers of fermion operators at particular instants of time.
Some correlations of groups of even numbers of operators at a given time
may become indirectly affected, but this type of
correlations feels only quantum statistics, which controls the
occupation of magnetic states $|\uparrow,0>, |0,\downarrow>$ in 
correspondence with the nonmagnetic ones $|\uparrow,\downarrow>, |0,0>$
and remains static until a noncommuting term such as a transverse field or 
a fermion hopping term etcetera is added to the Ising Hamiltonian.
As for the fermionic Ising spin glass, the
grand canonical description with a chemical potential $\mu$ and given by 
\begin{equation} 
{\cal{K}} = H-\mu N = - \sum_{ij}J_{ij}\sigma_i \sigma_j - 
\sum_ih_i\sigma_i-\mu \sum_{i,\sigma} n_{i,\sigma} 
\end{equation} 
appears standard, when finally the condition of independent and gaussian 
distributed random bonds $J_{ij}$ is imposed. The physics of this model 
is however far more complicated firstly than the 2--state parent model on 
spin space and secondly than related classical spin 1 models,
like the BEG--model \cite{BEG}, or the diluted 2--state per site model. 
Complications that arise in understanding the phase diagram of the are
deferred to the subsequent paper II.\\ 
According to the statement above, the SK--model and this model prove to 
be very different even at half--filling.\\
Expressing the partition function with the help of Grassman integrals the 
disorder averaging of the free energy is performed by means of the 
replica trick $\beta [F]_{av}= [\log Z]_{av}=\lim_{n \to 0}\frac{1}{n} 
(1-[\Pi_{\alpha =1}^n Z^{\alpha}]_{av})$.\\ 
The decoupling of eight--fermion correlations generated by the gaussian 
disorder average is rather standard and briefly described in the 
Appendix.\\
The fermionic standard model can be easily generalized or transformed to 
match more complicated and realistic systems.\\
While this paper is mainly devoted to the 
quantum--dynamical representation of Parisi symmetry  
in the field theory of the fermionic Ising spin glass, earlier studies of
spin correlations in more complicated models like metallic spin glasses also 
exist \cite{ROMB,sro}, and the analysis of fermionic correlations subject to 
RPSB promises many new results
in the light of the present results for the insulating cases.
\\ 
Writing a generalized grand canonical hamiltonian
\begin{equation}
{\cal{K}}=\sum_{ij}t_{ij}a^{\dagger}_{i\sigma}a_{j\sigma}
-\sum_{ij}J_{ij}^{\alpha}\sigma_i^{\alpha}\sigma_j^{\alpha}+
\left[ \left[U\right]_{av}+\delta U\right]\sum_i 
n_{i\uparrow}n_{i\downarrow}
+\sum_{i,j;\sigma,\sigma'}V(i-j)n_{i\sigma}n_{j\sigma'}
-\sum_{i,\alpha}h_{i}^{\alpha}\sigma_i^{\alpha}
-\mu\sum_{i,\sigma} n_{i\sigma}  
\end{equation} 
one comprises several subclasses such as itinerant magnetism (with 
emphasis on itinerant spin glasses), 
the Hubbard model including effects of disorder in U for 
example. Thus the general model takes care of Coulombic effects, both due 
the Hubbard interaction with and without random fluctuations of the coupling U,
and due to the long--range bare Coulomb repulsion V(i-j), and secondly 
allows for transport through either random or nonrandom fermion hopping. 
While the Hubbard interaction for example controls the crossover from 
4--state per site to an 3--state per site space at $U=\infty$ and 
adds a second magnetic interaction (without frustration if not due to a 
lattice), the long--range part of the Coulomb interaction is also important
due to its power to affect or even create gaps for example \cite{efros}.
Generalization due to longitudinal or transverse fields and due to 
Heisenberg type of interactions are rather standard and not of main 
interest. Both the metallic Ising spin glass (part 1 and 2) and the 
Hubbard model are known to be extremely hard problems even in the mean 
field limits and due to O(1) role of their corresponding Onsager reaction 
fields. On the other hand it is just this common point which, beyond the
physical relevance of Coulomb effects in fermionic spin glasses, calls for
a joint treatment. It is the $d=\infty$--method of the Hubbard model (called
dynamic mean field theory DMFT \cite{vollhardt} or local impurity 
selfconsistent approximation LISA \cite{rozenberg} by their inventors),
which we want to include to some extent and compare with the spin glass
results. One of the interesting questions concerns the type of competition that 
arises in the limit of vanishing $\overline{U}$, which implies an equal 
weight for local pairing states and for magnetic states. 
The locality of the responsible interaction renders
this problem different from frustration in spin glass problems. \\
A third group of fermionic spin glass models is given by the 
Kondo--coupling of
itinerant carriers to localized and randomly frozen spins. Since these spins
can be represented by pseudofermions coupled to a heat bath by imaginary
chemical potential, this model falls into the class of interacting systems
with two fermion species.  These s--d models, which apply as well to 
CuMn as to the
doped II--VI semiconductor CdMnTe \cite{terrypenney,chud}, and to heavy
fermion systems \cite{georges,menovsky,mentink,steglich,Maple,Loehn} are 
written in tight binding form by
\begin{equation} 
K=-\sum_{ij}t_{ij}a_{i\sigma}^{\dagger}a_{j\sigma} - \sum_i
a^{\dagger}_{i\sigma}\sigma^{\alpha}_{\sigma\sigma^{\prime}}
a_{i\sigma^{\prime}}s_i- \sum_{ij}J_{ij}s_i s_j -\frac{i}{2}\pi
T\sum_{i\lambda}d^{\dagger}_{i\lambda}d_{i\lambda}
-\mu\sum_{i\sigma}a^{\dagger}_{i\sigma}a_{i\sigma}
\end{equation} 
with the localized spins s given in terms of fermion
operators $d$ by $s_i\equiv d^{\dagger}_{i\lambda}
\sigma^{\alpha^{\prime}}_{\lambda\lambda^{\prime}} d_{i\lambda^{\prime}}$
where the simplest imaginary chemical potential $-i\frac{\pi}{2} T$ given 
here models the Hilbert space constraint for spin quantum number 
$\frac{1}{2}$. 
Higher spin quantum numbers can be treated with a little more effort
\cite{popfed,veits}. In the present paper we assume that the localized spins
are integrated out and end up in a single species model of a type 
described above. We do not claim that all physical phenomena of Kondo systems
can be described in this way, but the extraordinary difficulties 
encountered in theory
of fermionic spin glasses requires at present the limitation to the simplest
models. On the other hand the results given here for the fermionic Ising 
spin glass show that a quantum Parisi phase is to be expected in general.

\section{Basic elements and Symmetries of the Many Body Theory}
Conventional many body theory usually discards effects of statistical
fluctuations of interactions such as exchange-- or Hubbard--interactions.
This type of randomness is however not a rare event; a typical example
in the case of substitutional disorder is given by the so--called randomly
placed U--centers, a picture used by Hertz \cite{Hertz76} in an early and
pioneering diagrammatic approach to itinerant spin glasses.\\
Random interactions can lead to unique and decisive physical properties
in all sorts of fermionic quantum spin glasses.
The entire class of these systems cannot be exhaustively described on the 
spin
level as the example of effects of complex magnetic order on
low energy excitations and transport behaviour showed \cite{epl}.\\
In this chapter we provide details on technicalities and on physical results
obtained by the use of the elements of this new type of many body theory.\\
One of the major novelties stems from the necessity to include broken replica
permutation symmetry of Parisi--type \cite{parisi79,parisi1,parisi2} in the
fermionic many body theory \cite{epl,roamg}.
We shall demonstrate that the long--time behaviour of averaged
fermion propagators is determined by the highly nontrivial Parisi scheme.
This bears an interesting relation with the time--dependent formalism
intended to avoid the replica trick and applied to classical spin glasses
with additional Glauber dynamics. Our present theory however maintains the
replica trick together with quantum dynamics, which is enforced by the
Hamiltonian rather than being added due to the effect of external degrees of
freedom which provide a heat bath.
\subsection{Green's functions and generating functionals for fermionic spin
glasses}
The fermionic Green's function, denoted by ${\cal
G}_{i,\sigma}(\epsilon_l)$, and whose standard definition in terms of fermion
operators on a lattice reads
\begin{equation}
{\cal{G}}_{ij}(\tau,\tau')=-\left[<T_{\tau}(a_{i\sigma}(\tau)
a^{\dagger}_{j\sigma'}(\tau'))>\right]_{av}
\end{equation}
is studied in detail in this paper. It involves apart from the usual
quantum statistical average $<...>$ an additional disorder average which
uses probability distributions $P(J_{ij})$ for spin glasses or $P(U)$
for a random U Hubbard interaction for example.
In order t be able to study a Parisi type quantum solution, we choose the
replica formalism to cope with these disorder averages.
This formalism builts into fermion field theory the technicalities necessary
to deal with the typical problems that arise from randomness present in
many body interactions. Those are first the presence of at least
8--fermion correlations in the replicated action and secondly the broken
replica symmetry in connection with glassy order.

The method of time slicing allows to start out from a representation of   
the fermion propagator in terms of derivatives of a generating
functional with respect to anticommuting, generating (replicated) fields
$\eta,\overline{\eta}$ as
\begin{equation}
{\cal G}_i(\epsilon_l)=\frac{\delta}{\delta \eta^{a,l}_{i,\sigma}}
\frac{\delta}{\delta \overline{\eta}^{a,l}_{i,\sigma}}ln\prod\int d\psi
d\overline{\psi}exp[{\cal{A}}(\psi,\overline{\psi})+\eta\overline{\psi}-
\overline{\eta}\psi]|_{\eta=0,\overline{\eta}=0} ,   
\end{equation}
where ${\cal{A}}$ denotes the effective Grassmannian action. This action is
exactly of eigth order in the Grassmann fields, since we have chosen a 
complete
gaussian distribution of statistical fluctuations in two--body 
interactions.\\
a) {\it The fermionic Ising spin glass}
We start with the fermionic Ising spin glass $ISG_f$ as our first example,
thus avoiding complications from quantum--spin dynamics present for 
example in
the fermionic Heisenberg spin glasses or in metallic spin glasses of any 
kind.
Quantum spin dynamics, originating from (fermion) spin operators that
do not commute with the Hamiltonian, are interesting in itself but avoided
here to clarify the role of fermion quantum dynamics.
The latter is shown to be determined on the long time scale by the 
ultrametric
structure of the Parisi solution \cite{epl} so far only known to be 
present 
in (static) spin overlap order parameters.
We believe that this manifestation of ultrametricity in the $T=0$ quantum 
field
theory of frustrated interacting systems could also indicate a route to the
(to our best knowledge) open question of Parisi RPSB in the theory of
fermion localization. This refers to the standard part of the theory, whereas
the question of Anderson localization caused by disorder of frustrated
interactions is new and arises at a different level; a treatment independent
of the precise knowledge of the phase diagram and from the magnetic 
transition
would be hazardous. It is
clear that this expected phenomenon may be different from Anderson 
localization
of noninteracting systems as well as from Anderson--Mott transitions.\\
The model of gaussian distributed interaction constants allows to 
decouple the
effective eigth order fermion correlations by Q--fields, whose average is
static for models without quantum spin dynamics. Fluctuation fields
\begin{equation}
\delta Q^{ab}_i(t,t')\equiv Q^{ab}_i(t,t')-<Q^{ab}>
\end{equation}
are totally irrelevant in this case.\\
Taking care of an additional Coulomb interaction (including the local
Hubbard part) this situation remains unchanged.
This will be included below following the
effects of a nonrandom Hubbard interaction that plays an interesting role 
even in the absence of a fermion hopping hamiltonian. The cooperation with
the nonlocal spin interaction and the comparison with recent large d 
theories 
of the Hubbard model will then be of particular interest.
\subsection{Replicated Grassmannian action of the models with gaussian 
distributed interaction constants}
All models listed in section 2 can be represented by the use of anticommuting
fields. The partition function reads
\begin{equation}
Z=\int{\cal D}\Psi_{\gamma}(\tau)e^{-{\cal A}},
\end{equation}
where $\Psi(\tau)$ denotes a master Grassmann fields with imaginary time $\tau$
generated by time--slicing; this field 
$\Psi$ contains all Grassmann variables that may carry different 
components in order to take care of different fermion species. Its conjugate 
$\overline{\Psi}$ is related to $\Psi$ by charge conjugation.\\
The action can thus be written as
\begin{eqnarray}
{\cal A}&=& \int_0^{\beta\hbar}d\tau [\overline{\psi}_{\gamma\tau}
[\partial_{\tau}-\mu_{\gamma}/\hbar]\psi_{\gamma\tau}
+\overline{\chi}_{\gamma\tau}[\partial_{\tau}-
\frac{i\pi}{2}T]\chi_{\gamma\tau}\nonumber\\
&+& {\cal H}(\overline{\psi}_{\gamma\tau},\psi_{\gamma\tau},
\overline{\chi}_{\gamma\tau},\chi_{\gamma\tau})]
\end{eqnarray}
where all $\psi$ and $\chi$ variables anticommute with each other.
The nonhermitean part, which serves to reduce the 4--state per site Fock space
to a spin $\frac{1}{2}$ space, can be absorbed by U(1) transformation into 
the fields
\begin{equation} 
\tilde{\chi}_{\gamma}(\tau)\equiv 
exp(i\frac{\pi}{2}\frac{\tau}{\beta\hbar})\chi_{\gamma}(\tau). 
\end{equation}
In this way the
periodicity of the new anticommuting field $\tilde{\chi}$ becomes
\begin{equation}
\tilde{\chi}(\tau=\beta)=-i\tilde{\chi}(\tau=0)
\end{equation}
a periodicity that induces semionic imaginary frequencies of the Fourier 
transformed fields $\tilde{\chi(\epsilon_l)}$ with $\epsilon_l\equiv
(2l+\frac{1}{2})\pi k_BT/\hbar$.\\

The frustrated magnetic interaction of the 
fermionic Ising spin glass contributes 
\begin{equation}
{\cal{A}}_{eff}^{(int)}=-\sum_{a,b}\sum_{i,j}\frac{M_J(r_i-r_j)}{\hbar^2}
\int_0^{\beta\hbar}d\tau
\int_0^{\beta\hbar}d\tau'\sigma_i^a(\tau)\sigma_i^b(\tau')\sigma_j^a(\tau)
\sigma_j^b(\tau')
\end{equation}
to the effective interaction, provided the distribution of the 
$J_{ij}\hspace{.01cm}'s$
is gaussian with 2nd moment $M(r_i-r_j)$. The spin fields are given in 
terms of the anticommuting fields by
\begin{equation}
\sigma^a_i(\tau)=\sum_{\lambda=\pm1}
\overline{\psi}^a_{i\lambda}(\tau)\lambda\psi^a_{i\lambda}(\tau)
\end{equation}

\section{Results obtained from Replica Symmetric Approximations}
\subsection{Generating functional for the fermionic 
Ising spin glass}
The derivation of the Green's function is first given in replica symmetric 
formalism in order to provide the simplest possible and transparent 
introduction. The replica--symmetric results are stable 
solutions only above the Almeida--Thouless line \cite{AT,roamg} and 
unstable below. 
The sole presence of random fluctuations in a many body interaction 
already complicates the form of fermion propagators and hence deserves 
attention before the ordered phase and Parisi RPSB in addition will be 
described.\\ 
We will hence ignore until the next section all sorts of replica symmetry 
breaking, ie the most important Parisi RPSB, also vector replica symmetry 
breaking (RVSB), recently brought up by Dotsenko and M\'ezard 
\cite{dotsmez} 
in the different context of classical random field systems, which would mean
that different copies of the averaged one fermion propagator have to be 
studied. Another type of RPSB could as well arise in form of a nonzero 
fermion
propagator between between different replicas.\\
The replica symmetric saddle point solution of the Q--field
\begin{equation}
Q^{ab}=q \quad {\rm for}\quad  a\neq b, {\rm and} \quad Q^{aa}=\tilde{q}  
\end{equation}
allows to find the Green's function in the form
\begin{equation}
{\cal{G}}^a_{i,\sigma}(\epsilon_l)=\frac{\delta}{\delta\overline{\eta}_{\sigma}
^{a,l}}\frac{\delta}{\delta\eta^{a,l}_{\sigma}}ln \int_z^G\prod_a\int_{y_a}^G
\prod_a e^{\Phi_a(\eta^a,\overline{\eta}^a)}|_{_{\eta=\overline{\eta}=0}},
\end{equation}
where $y^a$ and $z$ denote replica--local and replica--global 
spin--decoupling fields and the gaussian integrations over these fields 
are written in the abbreviated notation 
$\int_{x}^G\equiv\int_{-\infty}^{\infty}\frac{dx}{\sqrt{2\pi}}
e^{-\frac{1}{2}x^2}$. 
The global field z decouples
$(\sum_a\overline{\psi}_{\sigma}^a\sigma\psi_{\sigma}^a)^2$ and 
$y^a$--integration takes 
care of VRSB quite naturally within the grassmannian formalism.\\
The functional can now be expressed in terms of the Green's function operator
g by
\begin{eqnarray}
exp\left[\sum_a\Phi_a(\eta,\overline{\eta})\right]
&=&\prod\int d\psi d\overline{\psi}
e^{\sum_{a,l,\sigma}\overline{\psi}^{a,l}_{\sigma}g^{-1}_{l,\sigma}(z,y^a)
\psi^{a,l}_{\sigma} + \eta^{a,l}_{\sigma}\overline{\psi}^{a,l}_{\sigma}
-\overline{\eta}_{\sigma}^{a,l}\psi_{\sigma}^{a,l}} \nonumber\\
&=& exp\left[\sum_a\Phi_{a}(0,0)\right]
exp\left[-\sum_{a,l,\sigma}g_{l\sigma}(z,y^a)
\overline{\eta}_{\sigma}^{a,l}\eta_{\sigma}^{a,l}\right],
\end{eqnarray}
with
\begin{eqnarray}
g_{l,\sigma}(z,y^a)&\equiv&g_{\sigma}(\epsilon_l|z,y^a)\nonumber\\
&=&\frac{1}{i\epsilon_l+\mu+\sigma(h + J\sqrt{q} z + J\sqrt{\tilde{q}-q}
y^a)}.
\end{eqnarray}
For brevity we shall use below the effective field
$\tilde{H}(z,y^a)=h+J\sqrt{q}\hspace{.1cm}z+J\sqrt{\tilde{q}-q}\hspace{.1cm}
y^a$.
Using the above equations the Green's function assumes the form
\begin{equation}
{\cal G}_{\sigma}^a(\epsilon_l)=\frac{\int_z^G\prod_{_{a^{\prime}}}
\int_{_{y^{a\prime}}}^G g_{\sigma}(\epsilon_l|z,y^a)
exp(\sum_{a^{\prime}}\Phi_{a^{\prime}}(z,y^{a^{\prime}}|0,0))}{\int_z^G
\prod_{_{a^{\prime}}}\int_{_{y^{a\prime}}}^G
exp(\sum_{a^{\prime}}\Phi_{a^{\prime}}(z,y^{a^{\prime}}|0,0))},
\end{equation}
which turns into
\begin{equation}
{\cal G}_{\sigma}(\epsilon_l)=
\int_z^G\frac{\int_y^G g_{\sigma}(\epsilon_l|z,y)exp(\Phi(z,y|0,0))}{\int_y^G
exp(\Phi(z,y|0,0))}
\end{equation}  
in the replica limit.\\
The rs--functional $\Phi(z,y^a)\equiv \Phi(z,y^a|0,0)$ is regularized at
$\tilde{H}=0$ and $\mu=0$ in order to retain any physically relevant
dependence. This results in
\begin{equation}
\Phi(z,y)=\sum_{\epsilon_l}\left[ln((\epsilon-i\mu)^2+\tilde{H}^2(z,y))-
ln\hspace{.1cm} \epsilon_l^2\right]
\end{equation}
The frequency sum evaluated in \cite{roamg}
yielded 
$e^{\Phi(z,y)}=2\left[cosh(\beta\mu)+cosh(\beta\tilde{H}(z,y))\right]$
which, together with the result
$\int_y^G e^{\Phi(z,y)} = 2\left[cosh(\beta\mu)+cosh(\beta\tilde{H}(z,0))
e^{\frac{1}{2}\beta^2 J^2(\tilde{q}-q)}\right]$,
shows that the replica--symmetric fermion Green's function follows from
\begin{equation}
{\cal{G}}_{\sigma}(\epsilon_l)=\int_z^G\int_y^G\frac{cosh(\beta\mu)+
cosh(\beta\tilde{H}(z,y))}{cosh(\beta\mu)+
cosh(\beta\tilde{H}(z,0))e^{\frac{1}{2}\beta^2 
J^2(\tilde{q}-q)}}g_{\sigma}(\epsilon_l|z,y) 
\label{20}
\end{equation}
\subsubsection{Fermion Green's Function in the disordered phase ($T>T_f$)}
The fermionic Ising spin glass does not allow for spin dynamics unless
Glauber dynamics introduces it by coupling to a heat bath. This would   
be analogous to the SK--model. The $ISG_f$'s Green's function however is
always dynamic and the way in which dynamical behaviour shows up in the
spin glass is however nontrivial. The simplest task is to determine first
the Greens function above the freezing temperature and to compare it for
example with that of a random Hubbard model in the local limit.
The fermionic Ising spin glass allows for an exact evaluation of the bare
fermion propagator
${\cal{G}}_{ij,\sigma}=[<\frac{1}{i\epsilon_n + \mu + \tilde{H}}>]_{av}$
in the disordered phase;
here $\tilde{H}$ denotes the usual effective field and the
double average refers to the replica--local and the
Parisi block decoupling fields. The result can be written in the form   
\begin{eqnarray}
& &{\cal{G}}_{ij,\sigma}(\epsilon_n)=\label{21}\\
&-&i\sqrt{\frac{\pi}{2J^2\tilde{q}}}
\sum_{\lambda =0,\pm 1} \frac{(2-\lambda^2)e^{\frac{1}{2}(\beta
J\lambda)^2\tilde{q}}ch((1-\lambda^2)\beta\mu_{\sigma})}{exp(\frac{1}{2}\beta^2
J^2\tilde{q})+ch(\beta\mu_{\sigma})}
(1-Erf(\frac{\epsilon_n-i\mu_{\sigma}+i\lambda\beta
J^2 \tilde{q}}{\sqrt{2\tilde{q}}J}))
e^{(\frac{\epsilon_n-i\mu_{\sigma}+i\lambda\beta
J^2\tilde{q}}{\sqrt{2\tilde{q}}J})^2} \delta_{ij},\nonumber
\end{eqnarray}
where $\mu_{\sigma}\equiv \mu +\sigma H$ includes a magnetic field $H$ and 
$\epsilon_n=(2n+1)\pi k_B T/\hbar$.
By analytical continuation to real energies (the error function with
imaginary argument may then be expressed by the confluent hypergeometric
function with real argument) one
easily extracts the disorder averaged electronic density of states 
\cite{brro}
\begin{equation}
\rho_{\sigma}(\epsilon)=-\frac{1}{\pi}\hspace{.1cm}{\cal{I}}mG^R_{\sigma}(E)=
\frac{1}{\sqrt{2\pi\tilde{q}}J}e^{-\frac{(\epsilon+\mu)^2}
{2J^2\tilde{q}}}
\frac{ch(\beta\mu)+ch(\beta(\epsilon+\mu))}{ch(\beta\mu)+ch(\beta
H)exp(\frac{1}{2}\beta^2 J^2\tilde{q})}
\end{equation}
and the real part of the Green's function as
\begin{eqnarray}
{\cal{R}}e 
G_{\sigma}(E)&=&\frac{ch(\beta\mu_{\sigma})}{ch(\beta\mu_{\sigma})
+e^{\frac{1}{2}
\beta^2 
J^2\tilde{q}}}\frac{E}{J^2\tilde{q}}\hspace{.01cm}_1F_1(\frac{1}{2},\frac{3}{2};
\frac{E^2}{2 J^2\tilde{q}})\nonumber\\
&+&\frac{exp(\frac{1}{2}\beta^2 J^2\tilde{q}-\frac{(E+\beta 
J^2\tilde{q})^2}{2 J^2\tilde{q}})}{ch(\beta\mu_{\sigma})
+exp(\frac{1}{2}\beta^2 J^2\tilde{q})}\frac{E+\beta J^2\tilde{q}}{2 
J^2\tilde{q}}\hspace{.01cm}_1F_1(\frac{1}{2},\frac{3}{2};\frac{(E+\beta 
J^2\tilde{q})^2}{2 J^2\tilde{q}})
\nonumber\\
&+&\frac{exp(\frac{1}{2}\beta^2 J^2\tilde{q}-\frac{(E-\beta
J^2\tilde{q})^2}{2 
J^2\tilde{q}})}{ch(\beta\mu_{\sigma})+exp(\frac{1}{2}\beta^2 
J^2\tilde{q})}\frac{E-\beta J^2\tilde{q}}{2 
J^2\tilde{q}}\hspace{.1cm}_1F_1(\frac{1}{2},
\frac{3}{2};\frac{(E-\beta J^2\tilde{q})^2}{2 J^2\tilde{q}})
\end{eqnarray}
where 1F1 denotes the confluent hypergeometric function. \\ 
\subsubsection{Spin-- and charge--response function derived from the
spin glass fermion propagator}
As a basis for calculations in the spin glass phase, and in particular
under respect of replica symmetry broken fermion propagators, we
demonstrate the easiest example of spin-- and charge--susceptibilities
above the freezing temperature. Both correlations given by
\begin{eqnarray}
\chi_{\zeta}&\equiv& 
T^2[\sum_{\sigma,\epsilon}g^2_{\sigma}(\epsilon_l)]_{\Phi(y)}-
[(T\sum\sigma^{\zeta}g_{\sigma}(\epsilon_l))^2]_{\Phi(y)}\\
&=&T^2\sum_{\sigma,\epsilon}{\cal{G}}^2_{\sigma}(\epsilon_l)
-(T\sum_{\sigma\epsilon}\sigma^{\zeta}{\cal{G}}(\epsilon))^2
\end{eqnarray}
are static and satisfy the relation $\chi_{\sigma}+\chi_{\rho}=2\nu\equiv 
2[n]$.
\subsubsection{Fermion Green's function in the spin glass ordered phase 
($T<T_f$) (replica--symmetric approximation)}
\begin{figure}
\vspace{1cm}\hspace{1.5cm}
\epsfig{file=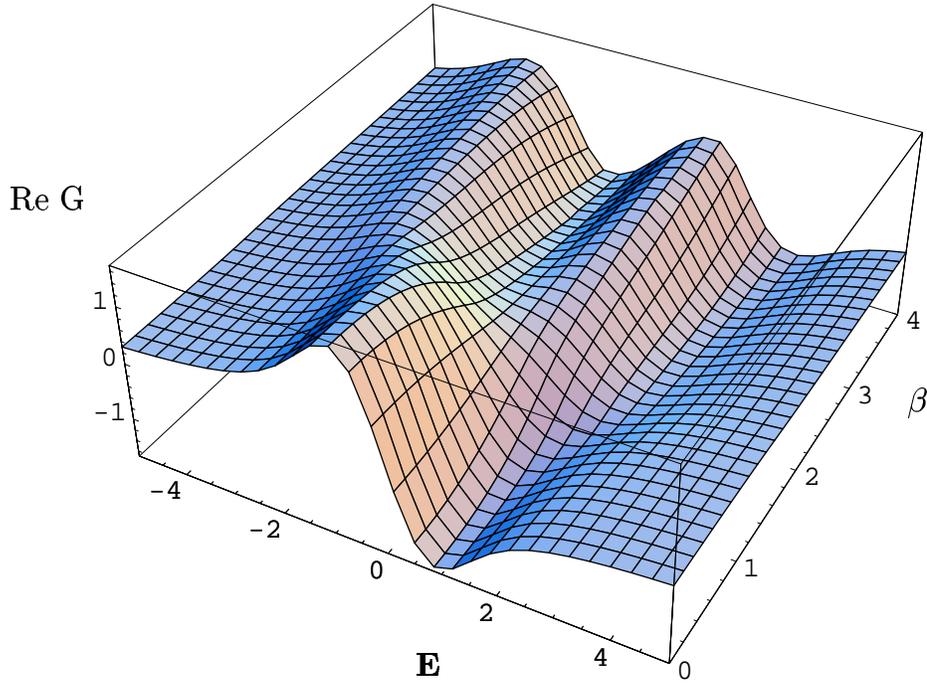,width=10cm,angle=270}
\caption{The real part of the replica--symmetric fermion
Green's function above and below the freezing temperature $.6767 J$
(-${\cal{R}}e[G]$ is shown) as a function of $\beta J=\frac{J}{T}$}
\label{f7a}
\end{figure}

\begin{figure}
\hspace{1cm}
\epsfig{file=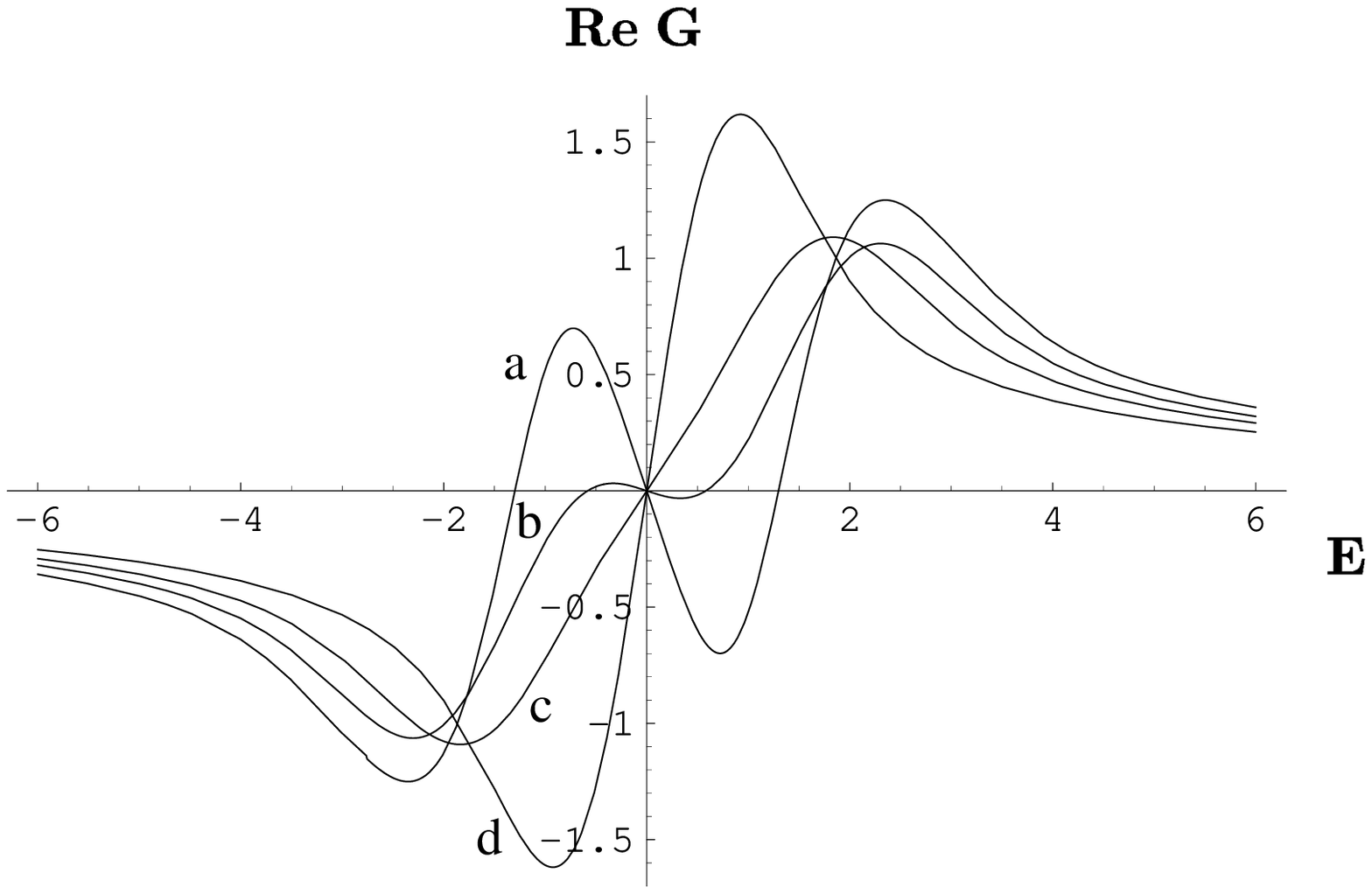,width=8cm,angle=270}
\caption{Cross sections through preceding figure (rotated) of Re[G] at 
temperatures
$T=\infty$ (curve a), $T=T_f$\hspace{.1cm}(b), $T=.5 J$\hspace{.1cm}(c), 
and $T=.2 J$\hspace{.1cm}(d)}
\vspace{2.5cm}\hspace{1.5cm}
\epsfig{file=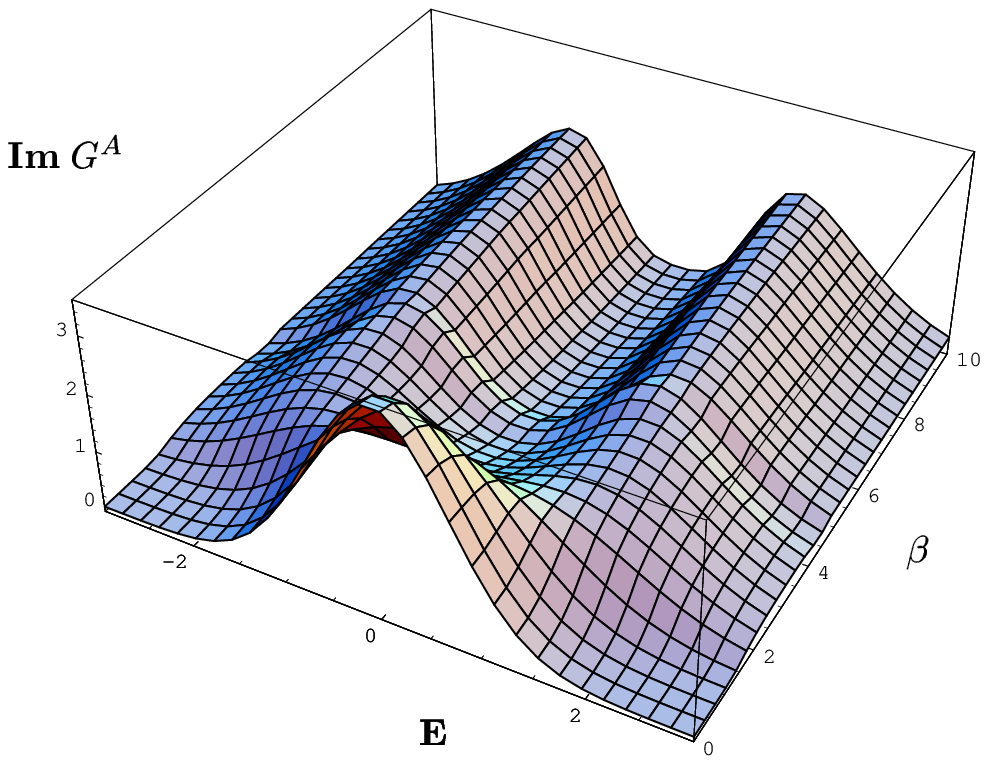,width=8cm,angle=270}
\caption{The imaginary part of the replica--symmetric $G^A$ ($\sim DoS$)
as a function of energy $E=\epsilon$ ($\mu=0$) and $\beta J=\frac{J}{T}$.}
\label{f7b}
\end{figure}

\begin{figure}
\hspace{2cm}
\epsfig{file=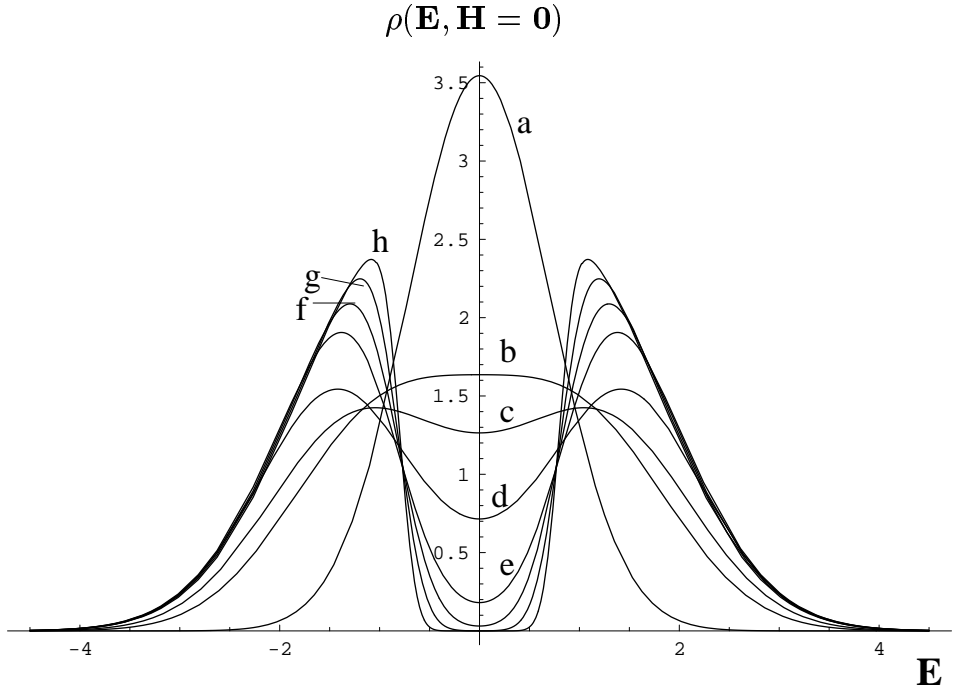,width=10cm,angle=270}
\caption{Creation of the spin glass related gap as the temperature 
falls below $T_f$ (cross sections of preceding figure): curve 
a: $T=\infty$, b: $T\approx T_f$, 
c: $T=J/2$, d: $T=J/3$, 
f: $T=J/6$, $T=.1J$, g: $T=.05J$ , and h: $T=.025 J$.} 
\label{f7c}
\end{figure}

Despite the fact that the replica symmetric result is unstable below the
fermionic Almeida Thouless line, hence in any case below the freezing 
temperature, the fermion propagator is complicated to an extent that requires
insight from its simplest realization. This is the extension of Eq.(\ref{21})
by including the effect of one replica symmetric spin glass order parameter q.
By evaluating Eq.(\ref{20}) we obtain on the imaginary frequencies
\begin{eqnarray}
{\cal{G}}^{(0)}_{\sigma}(\epsilon_l)&=&\frac{i}{4 
J}\frac{1}{\sqrt{\tilde{q}-q}}
\int_{-\infty}^{\infty}dz e^{-\frac{1}{2}z^2}
\frac{1}{ch(\beta\mu)+Exp(\frac{1}{2}\beta^2 
J^2(\tilde{q}-q))ch(\beta J\sqrt{q}z)} \nonumber\\
& &\hspace{-.2cm}[2 
ch(\beta\mu)e^{\frac{(\epsilon_l-i(\mu+J\sqrt{q}z))^2}{2J(\tilde{q}-q)}}
(1-Erf(\frac{\epsilon_l-i(\mu+J\sqrt{q}z)}{J\sqrt{2(\tilde{q}-q)}}))\nonumber\\
& &\hspace{-1cm}+ \sum_{\lambda=\pm 1}e^{\lambda\beta J\sqrt{q}z
+\frac{1}{2}\beta^2 J^2(\tilde{q}-q)}
e^{(\epsilon_l-i(\mu+J\sqrt{q}z+\lambda\beta 
J(\tilde{q}-q))^2/(2 J(\tilde{q}-q))}
[1-Erf(\frac{\epsilon_l-i(\mu+J\sqrt{q}z+\lambda\beta J(\tilde{q}-q)}{
J\sqrt{2(\tilde{q}-q)}})]]
\label{24}
\end{eqnarray}
The analytic continuation of this expression provides us with the replica 
symmetric density of states, showing a magnetic hardgap as discussed below,
and the real part of $G^{(0)}(E)$ given by
\begin{eqnarray}
Re[G^{(0)}(E)]&=&\frac{1}{\sqrt{2\pi}J(\tilde{q}-q)}\int_{-\infty}^{\infty}dz
Exp(-\frac{1}{2}z^2)\frac{1}{ch(\beta\mu)
+Exp(\frac{1}{2}\beta^2 J^2(\tilde{q}-q))ch(\beta J\sqrt{q}z)}\label{25}\\
& &\hspace{-2cm}
\{[ch(\beta\mu)Exp(-\frac{(E+J\sqrt{q}z)^2}{2J^2(\tilde{q}-q)})
(E+J\sqrt{q}z)_1F_1(\frac{1}{2},\frac{3}{2};\frac{(E+J\sqrt{q}z)^2}{2
J^2(\tilde{q}-q)})
+\sum_{\lambda=\pm 1}(E+J\sqrt{q}z+\lambda\beta J(\tilde{q}-q))\nonumber\\
& &\hspace{-2.5cm}Exp\left[\lambda\beta 
J\sqrt{q}z+\frac{1}{2}\beta^2 J^2(\tilde{q}-q)-
\frac{(E+J\sqrt{q}z+\lambda\beta J(\tilde{q}-q))^2}{2 
J^2(\tilde{q}-q)}\right]
\hspace{.1cm}_1F_1(\frac{1}{2},\frac{3}{2};\frac{(E+J\sqrt{q}z
+\lambda\beta J(\tilde{q}-q))^2}{2 J^2(\tilde{q}-q)})) \nonumber
\end{eqnarray}
The asymptotic behaviour either for large energies or for $T\rightarrow 0$
can be checked by the use of $1-Erf(x)\cong 
\frac{1}{\sqrt{\pi}}\frac{exp(-x^2)}{x}$ and the functional relation
$Erf(x)=\frac{2 
x}{\sqrt{\pi}}\hspace{.01cm}_1F_1(\frac{1}{2},\frac{3}{2},-x^2)$
\cite{GradRyz}.\\
The result Eq.(\ref{25}) is displayed in Fig.(\ref{f7a}). 
Comparison with the extrapolation of the (paramagnetic) $q=0$--result
below $T_c$, the role of the (replica--symmetric) susceptibility as a 
measure of the spread of just two wells forming below the freezing 
temperature is shown. 
This spread is seen to be related to the size of the magnetic hardgap in this 
approximation. We shall come back to the essential changes of the fermion
propagator in the small energy-- and in the long--time regime below in the
section on replica permutation symmetry breaking and Parisi symmetry.\\
The fermion density of states in replica symmetric approximation already
reveals the strong coupling between spin glass order and fermionic properties.
The strongest effect emerges in the zero temperature limit. 
After analytic continuation $i\epsilon_l\rightarrow\epsilon+i0$ 
we find
\begin{equation}
\rho_{\sigma}(\epsilon)=
\frac{cosh(\beta\mu)+cosh(\beta E)}{2\pi J\sqrt{\tilde{q}-q}}
\int_{-\infty}^{\infty}dz\frac{e^{-\frac{1}{2}z^2-
\frac{1}{2}\frac{(H+J\sqrt{q}z+\sigma E)^2}{J^2(\tilde{q}-q)}}}
{cosh(\beta\mu)+cosh(\beta\tilde{H}(z,0))
e^{\frac{1}{2}\beta^2 J^2(\tilde{q}-q)}}
\label{dos}
\end{equation}
The $T\rightarrow 0$ limit is not simple. We first consider the regime
\begin{equation}
|\mu|<\frac{1}{2}\beta J^2(\tilde{q}-q).
\end{equation}
One should note that $\tilde{q}-q=O(T)$, whence this condition requires
the Fermi energy to satisfy
\begin{equation}
E_F<\frac{J}{\sqrt{2\pi}}.
\end{equation}
This value was obtained from the rs--selfconsistent order parameter equations
above. One may prefer to identify $\beta J(\tilde{q}-q)$ with the linear
susceptibility $\chi$ of the fermionic Ising spin glass $ISG_f$.
\begin{equation}
\rho_{\sigma}(\epsilon)=\frac{e^{-\frac{1}{2}\beta^2
J^2(\tilde{q}-q)}(cosh(\beta\mu)+cosh(\beta E))}{2\pi 
J\sqrt{\tilde{q}-q}}
\int_{-\infty}^{\infty}dz
\left[\frac{e^{-\frac{1}{2}z^2-\frac{1}{2}
\frac{(J\sqrt{q}z+H+\sigma 
E)^2}{J^2(\tilde{q}-q)}}}{cosh(\beta\tilde{H}(z,0))} +...\right]
\end{equation}
The $T\rightarrow 0$ limit is most easily obtained by integrating over the 
rescaled variable $u\equiv\beta\tilde{H}(z,0)$. This isolates the leading
contribution as
\begin{equation}
\rho_{\sigma}(\epsilon)=
\frac{cosh(\beta\mu)+cosh(\beta E)}{2\pi\beta
J^2\sqrt{q(\tilde{q}-q)}}e^{-\frac{1}{2}(\beta^2 J^2(\tilde{q}-q)
+\frac{H^2}{J^2 q}+\frac{E^2}{J^2(\tilde{q}-q)})} I(T,E),
\end{equation}
where the integral 
\begin{equation}
I(T,E)=\int_{-\infty}^{\infty}du
\frac{1+O(T u^2)}{cosh(u)e^{-\frac{TEu}{J^2(\tilde{q}-q)}}}
\end{equation}
converges for energies satisfying the condition
\begin{equation}
E<\beta J^2(\tilde{q}-q).
\end{equation}
Since the maximal $|E|$ within this regime coincides with the limit of a
vanishing prefactor and hence a vanishing density of states in the 
$T\rightarrow 0$ limit, this energy is to be identified with the gap edge.
We call the gap energy $E_g(h)$. The replica--symmetric solution for energies
inside the $T=0$--gap (and Fermi energies within half of this gap) is hence
\begin{equation}
E_g(H):=\sqrt{\frac{2}{\pi}} J e^{-\frac{1}{2}\frac{H^2}{J^2}}
\end{equation}
and the final result for the leading low temperature DoS--contribution,
which vanishes exponentially within the gap $|E|<E_g(H)$, reads
\begin{equation}
\rho_{\sigma}(\epsilon)=
\frac{cosh(\beta\mu)+cosh(\beta E)}{2J\sqrt{\beta E_g(H)}
cos(\frac{\pi}{2}\frac{E}{E_g(H)})}
e^{-\frac{1}{2}\frac{H^2}{J^2}+\frac{1}{2}(1-\frac{H^2}{J^2})
(\frac{E^2-E_g^2(H)}{J^2})+\frac{E^2+E_g^2(H)}{T E_g(H)}}
\end{equation}
Two striking aspects emerge:\\
i) there is neither a spin dependence of the gapwidth nor of the
leading and exponentially decaying low temperature density of states
despite the magnetic field dependences;\\
ii) the given result is valid only for $|\mu|<\frac{1}{2}E_g(H)$, while there
is no limitation on $|E|\equiv|\epsilon+\mu|$. Thus the conclusion on the
gapwidth is limited to a Fermi energy varying over half the gapwidth.\\
iii) one would like to see how the Fermi energy can be moved through a
gap edge in order to obtain a non half--filled ground state. 
From the analysis of the free energy we know however that phase separation
occurs beyond $|\mu_0|=\frac{1}{2}E_g(H)$, at least in the 
replica--symmetric approximation. Hence we believe that instanton
solutions may have to be used, replacing the homogeneous saddle point 
solutions above. 
We conjecture that an instability, identified as a second negative
eigenvalue of the Hessian matrix -- the first being just the known Parisi--RSB
instability --, indicates just this breakdown of the spatially constant 
solution in the regime of first order transition from a half--filled to a
completely filled (or empty respectively) system.\\
One should nevertheless realize that only $\mu=0$ corresponds to half--filling
at any temperature; moreover it is interesting to compare the domain of
continuous thermal spin glass transitions which extends to the tricritical
points at either $|\mu(T_{c3})|\approx .96125$ at $T_{c3}=\frac{2}{3}J$ with 
the $|\mu(T=0)|=\frac{J}{\sqrt{2\pi}}\approx 0.398942 J$. \\
Real and imaginary parts of the Green's function are plotted for various
cases in Figs.(\ref{f7a},\ref{f7b},\ref{f7c}).
\pagebreak
\subsubsection{The magnetic hardgap of the DoS at $T=0$}
The density of states $\rho(T=0,|E|>E_g(h))$ is nonzero; it can be evaluated
by the saddle point method. We start out from Eq.(\ref{dos}). The very same 
condition $|\mu|<\frac{1}{2}E_g(h)$
applies again. We shall come back to this below.\\
While $q, \chi$, and $\nu(\mu)$ can 
be obtained by extremizing the free energy, ie without explicit knowledge 
of the underlying density of states (DoS), this fermionic DoS     
is of course related to the magnetic properties of the system and its 
analysis requires the solution of several selfconsistency
equations for spin-- and for charge correlations. 
We obtained these solutions for low temperatures and for the 
entire range of magnetic fields and of single fermion excitation energies.
The existence of a hardgap of size $2E_g(H)$ is best appreciated in our
$T=0$--result for the density of states, which is given by
\begin{equation}
\rho_{\sigma}(\epsilon; H) = \frac{1}{\sqrt{2\pi}J}
\Theta(|E| - E_g(H))exp\left[-\frac{1}{2}\left[\frac{E}{J}
(1-\frac{E_g(H)}{|E|})+\sigma \frac{H}{J}\right]^2\right],
\label{2}
\end{equation}
with $E:=\epsilon+\mu$. For finite range models we expect exponentially
small corrections within the gap and due to Griffiths singularities.
Eq.(\ref{2}) is valid within the regime
$|\mu|<\frac{1}{2}E_g$, which corresponds to a half filled system at $T=0$
with filling factor
\begin{equation}
\nu=1+(1-q-T\chi)tanh(\beta\mu)  .
\end{equation}
The zero field ratio $E_g(0)/T_c$ increases with the chemical potential 
from 1.179 at $\mu=0$ to 1.238 at $\mu=J/\sqrt{2\pi}$.
In terms of $E_g(H)$ the following low temperature expansions 
(for linear susceptibility and order parameter), 
required for the exact evaluation of the DoS formula (\ref{dos}) at low T, 
were obtained as
\begin{eqnarray}
q &=& 1 - \frac{E_g(H)}{J}\frac{T}{J} + O(\frac{T^2}{J^2}),\\
\chi &=& \frac{E_g(H)}{J^2} + 
\frac{1}{2}\frac{E^2_g(H)}{J^3}(1-\frac{H^2}{J^2})\frac{T}{J} + 
O(\frac{T^2}{J^2}).
\end{eqnarray}  
These results show that i) the density of states is zero at $T=0$ in the
finite interval given by $|\epsilon+\mu|<E_g(H)$ and ii) the gapwidth 
shrinks as 
the magnetic field is increased. This will be the source of the negative 
magnetoresistance in the extended Ising spin glass model with charge transport 
as discussed below. The physical consequences of these spin glass related 
properties are hence in agreement with the experimental observations of 
crossover behavior in the low T resistivities mentioned above. 
For single fermion energies $E\equiv \epsilon+\mu$ smaller than the 
gap energy $E_g(H)$ and $|\mu| < \frac{1}{2}E_g(H)$
the density of states decays to zero exponentially as given by
\begin{equation}
\rho_{\sigma}(\epsilon)= \frac{1}{2\hspace{.1cm}J\hspace{.1cm}
cos(\frac{\pi}{2}\frac{E}{E_g(H)})}\hspace{.1cm} 
e^{-\frac{1}{4}\frac{E^2_g(H)-E^2}{J^2}(1-\frac{H^2}{J^2})
-\frac{1}{2}\frac{H^2}{J^2}}
(\beta E_g(H))^{-\frac{1}{2}}\hspace{.1cm}
\left[ch(\beta\mu)+ch(\beta E)\right]\hspace{.1cm}
e^{-\frac{1}{2}\beta\frac{E^2_g(H)+E^2}{E^2_g(H)}}.
\end{equation}
We find that the hardgap persists also in itinerant models with a 
fermion hopping term added to the nonconducting Ising Hamiltonian,
until the bandwidth exceeds a critical value.
The gap and its related properties of this extended itinerant spin glass 
model, whose magnetic phase diagram and --transitions have 
previously been analysed \cite{{sro},{brro}}, are discussed below.\\
\begin{figure}
\vspace{-2.5cm}\hspace{2cm}
\psfig{file=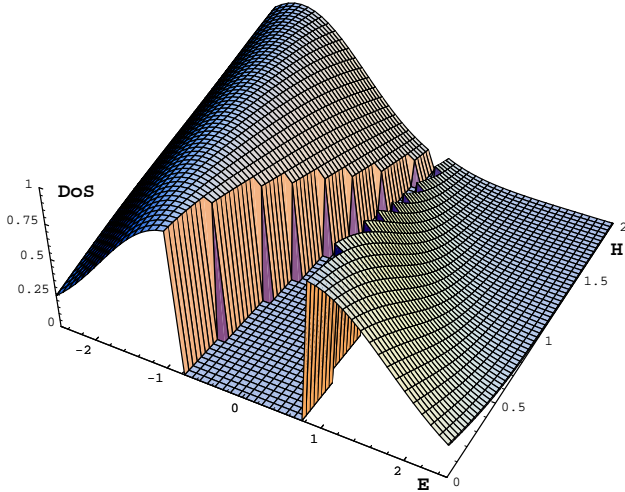,width=8.5cm,angle=0}
\vspace{-2.5cm}
\caption{Single particle density of states (DoS) for the fermionic Ising 
spin glass
($ISG_f$) versus energy and magnetic field in units of J. The hardgap 
centered around zero
energy is due to random magnetic correlations and manifests itself in an
activated behavior of the low temperature hopping conductivity of disordered
magnetic materials. The reduction of the gapwidth by an external 
magnetic field explains the observed negative magnetoresistance.}
\end{figure}      
Coulomb interaction effects are of diverse nature: the long range part
will still tend to depress the (remaining) DoS near the Fermi level and
thus stay particularly relevant when either
spin glass order is weak or absent, or (in the case of fully developed spin 
glass order) when the Fermi energy lies close to the gap edges.
The Hubbard coupling U leads to a shift of the chemical potential
($\mu\rightarrow \mu-\frac{U}{2}\nu$), and also to a shift 
$H\rightarrow H^{\prime}\equiv H+\frac{U}{2} m(H)$
of the applied magnetic field. 
This implies for finite temperatures that the Almeida Thouless line 
depends on the Hubbard coupling too.
\subsubsection{The replica--symmetric approximation of the $T=0$ 
fermion propagator}
The density of states could be used as the spectral weight in the 
Kramers--Kronig relation for the fermion Green's function. One may as
well extract the zero temperature limit of Eq.(\ref{24}). This results 
in the (bare $K=0$) fermion propagator ${\cal{G}}_{\sigma}^{(K=0)}(\epsilon)$
\begin{equation}
{\cal{G}}^{(0)}_{\sigma}(\epsilon)=-i\int_0^{\infty}\frac{dz}{\sqrt{2\pi}}
e^{-\frac{1}{2}z^2}\{\frac{1}{\epsilon-i(\mu+ 
z+\chi_0)}+\frac{1}{\epsilon-i(\mu-z-\chi_0)}\}
\label{40}
\end{equation}
with energies $\epsilon_l=(2l+1)\pi T\rightarrow \epsilon$ lying dense on 
the imaginary axis and $\chi_0=lim_{T\rightarrow 0}\beta(\tilde{q}-q)$ 
representing the RS--approximation of the susceptibility (here and in the 
following we set $J=1$ for brevity). The most remarkable fact being that the 
gaussian distribution of say 
self--energies $z+\chi_0$ are now cutoff, which is responsible for the 
hardgap described above. Recall that in 
Eq.(\ref{24}) one starts off from a complete gaussian average over all z.\\
Insertion the zero temperature results for the density of states, 
Eqs.(\ref{dos}) and (\ref{2}), in the spectral representation
\begin{equation}
{\cal{G}}_{\sigma}(\epsilon_l)=\int_{-\infty}^{\infty}d\epsilon'
\frac{\rho_{\sigma}(\epsilon')}{i\epsilon_l-\epsilon'}.
\label{spectral}
\end{equation}
confirms the representation of Eq.(\ref{40}).
The final integral in Eq.(\ref{40}) can be evaluated to give the following 
explicit solution 
\begin{equation}
{\cal{G}}^{(0)}_{\sigma}(\epsilon)=-\frac{1}{\sqrt{8\pi}}
\sum_{\lambda=\pm 1}\left[e^{-\frac{A_{\lambda}^2}{2}}
Ei(-\frac{A_{\lambda}^2}{2})+i\pi\lambda sign(\epsilon)
e^{\frac{A_{\lambda}^2}{2}}
\left[1-Erf(\frac{A_{\lambda}}{\sqrt{2}})\right]\right]
\end{equation}
with
\begin{equation}
A_{\lambda}(\epsilon)\equiv 
\epsilon-i\left[\mu+\lambda\chi_0\right], 
\end{equation}
and Ei(x) denoting the exponential integral function \cite{GradRyz}.
The results are given for zero magnetic field; the fact that the chemical 
potential appears in the form $\mu\pm\chi_0$ shows the effect of the Onsager 
reaction field. The oversimplification of the replica--symmetric 
approximation is revealed by the $\mu$--shift, resembling only 
contributions from up and down local fields.\\
The time--dependence is obtained from
\begin{eqnarray}
{\cal{G}}^{(0)}_{\sigma}(\tau)&=&\int_0^{\infty}\frac{dz}{\sqrt{2\pi}} 
e^{-\frac{1}{2}z^2} \{-\left[e^{-\mu_{+}(z)\tau}\theta(-\mu_{+}(z))+
e^{-\mu_{-}(z)\tau}\theta(-\mu_{-}(z))\right]\theta(\tau)\nonumber\\
&+&\left[e^{-\mu_{+}\tau}\theta(\mu_{+}(z))
+e^{-\mu_{-}\tau}\theta(\mu_{-}(z))\right]\theta(-\tau) \},
\end{eqnarray}
where we introduced effective chemical potentials
\begin{equation}
\mu_{\pm}(z)\equiv \mu\pm(z+\chi_0)
\end{equation}
that depend on the spin decoupling field and on the lowest order 
susceptibility. The latter represents the Onsager field contribution.\\
We recall that the given solution is valid for $\mu<\frac{1}{2}\chi_0$
(apart from its instability against RPSB that is taken into account 
below), which means half--filling at $T=0$.
Evaluation of the imaginary--time dependent propagator yields 
\begin{equation}
{\cal{G}}^{(0)}(\tau)=-\frac{1}{2}e^{-(\mu-\chi_0)\tau-\frac{\tau^2}{2}}
\left[1+Erf(\frac{\tau}{\sqrt{2}})\right]\theta(\tau)
+\frac{1}{2}e^{-(\mu+\chi_0)\tau
-\frac{\tau^2}{2}}\left[1-Erf(\frac{\tau}{\sqrt{2}})\right]\theta(-\tau)
\end{equation}
Alteration of this time--dedendent behaviour under RPSB will be an important
conclusion below. Before this problem is attacked, it is interesting 
to compare the replica symmetric fermionic spin glass propagator with a 
propagator of random U Hubbard models in the local limit. 
Fermion hopping, which introduces the full complexity of the Hubbard model, 
plays also an important role in promoting the insulating fermionic spin glass 
to a metallic spin glass model.\\
Since all of the complicated physics of both model classes is sandwiched 
between the (identical and solvable) fermion hopping model and the 
(different and solvable, provided an infinite--range magnetic interaction 
is considered) pure interaction limits, the latter limit offers the most 
interesting case for comparing the two solutions.   
\subsubsection{Comparison and mapping with the Green's function of a 
random U local Hubbard limit}
The wellknown exact solution of the local limit played an important 
role as starting points of recent many body theories of the Hubbard model
\cite{vollhardt,georges,Metzner}. The field--theoretic decoupling given 
by Vollhardt \cite{vollhardt} can be extended to account for disorder 
fluctuations in the coupling U. We shall compare the spin glass and the 
random U Hubbard model field theoretic techniques in Appendix B. For the
purpose of comparing the replica--symmetric fermion propagator of the 
fermionic spin glass with a potential random U Hubbard analog in the 
local limit, we do not need this apparatus. It is sufficient to consider 
a gaussian average over U with a properly chosen cutoff to see the similarity
with Eq.(\ref{40}).
As well one may compare the Green's function obtained above the freezing 
temperature with the local Hubbard limit.
At half--filling the Green's function reads
\begin{equation}
{\cal G}_{\sigma}(\epsilon_l)=\frac{1}{2}[\frac{1}{i\epsilon_l+U/2}
+\frac{1}{i\epsilon_l-U/2}]
\end{equation}
Since this is the exact solution for any half--filled U--realization in 
the local random U model, it is easy to evaluate not only gaussian 
distributions but also Lorentz--, box--, and 
semicircular--distributions of random U--couplings 
in order to obtain the disorder--averaged one particle Green's function. 
The average $<log Z>$ can also be derived. Thus these results serve also
as a interesting test field for the replicated field theory, which usually 
runs into formal difficulties whenever nongaussian distributions are 
involved.\\ 
For unrestricted gaussian--distributed U (equal weight of attractive and 
repulsive interaction) one obtains 
\begin{equation}
[{\cal G}_{\sigma}(\epsilon_l)]_{P_g(U)}=\pi 
e^{\epsilon_l^2/M_U}(1-Erf(\frac{|\epsilon_l|}{\sqrt{M_U}}))sgn(\epsilon_l)
\end{equation}
and the density of states is a simple gaussian
\begin{equation}
[\rho]_{P_g(U)}(\epsilon)=\sqrt{\frac{M_U}{8\pi}}e^{-2\epsilon^2/M_U}
\end{equation}
In order to generate a solution that matches the fermionic spin glass
propagator, Eq.(\ref{40}), one needs to cut off the gaussian distribution.\\
For $\mu=0$ it is just the Onsager reaction field \cite{Onsager} of the 
spin glass, which requires to introduce a finite average $<U>$ by
\begin{equation}
U=<U>+\delta U,
\end{equation}
where the fluctuations obey a gaussian distribution of nonnegative $\delta 
U$--fluctuations 
\begin{equation}
P^c_g(\delta U)=\frac{1}{\sqrt{2\pi}}exp(-\frac{1}{8}(\delta 
U)^2)\Theta(\delta U) 
\label{50}
\end{equation}
which maps the fermion propagator of 
this random U Hubbard interaction onto the replica--symmetric approximation
of the fermionic Ising spin glass propagator at $T=0$ and $\mu=0$, provided
(again given in dimensionless quantities)
\begin{equation}
<U> = 2\chi_0
\label{51}
\end{equation}
This mapping involves only repulsive but random interactions. Small 
random deviations from the average positive U are strongly weighted, which
renders the model perhaps a realistic one. The relation (\ref{51}) also 
provides an illustration in terms of the Hubbard interaction for the 
gap--width $2\chi_0$ of the fermionic spin glass.\\
It is interesting to imagine an additional fermion hopping term, thus a 
Hubbard model with the given U--fluctuations: realizations of large U can 
be described by a t--J model and the entire regime down to perturbative 
small U is contained with increasing weight. \\
The random U local Hubbard limit, introduced here for purely formal reasons,
is a zero--dimensional problem and one cannot expect to find a spin glass
transition like the one described for the Ising spin glass model
with nonlocal or even infinite--range interaction. Allowance for fermion
transport however reopens this possibility and requires further analysis,
also in relation with other metallic spin glasses.
Moreover, the problem of RPSB can be raised for the local limit of 
the random U Hubbard model too, which means that the exact solution in the
random case may not simple. We defer this analysis to another publication.\\
In contrast to the described situation, which contains a density of states 
gap identical to the fermionic spin glass by construction, a gaussian 
distribution without cutoff removes the gap and poses hence a 
problem to the existence of a Mott transition of the Hubbard model. 
A striking difference with
respect to the interaction gap of the frustrated nonlocal spin interaction
is to be noticed. In this respect the statistical fluctuations of the
Hubbard interaction compete with both the nonrandom part of U and with
the spin coupling $J_{ij}$. We first need to evaluate exactly the
partition function and the generating functional of the random local limit,
which passes through a time--dependent Grassmann field theory.\\
In contrast to fermionic spin glass models it is easy to consider 
distributions other than gaussian in the local Hubbard limit. The latter
may thus become helpful for the spin glass problem.\\
For example, the box distribution $P_b(U)$ yields the averaged Green's 
function 
\begin{equation}
[{\cal G}_{\sigma}(\epsilon_l)]_{P_b(U)}=\frac{-2i}{U_2-U_1}
\left[arctan\frac{U_2}{2\epsilon_l}
-arctan\frac{U_1}{2\epsilon_l}\right]
\end{equation}
which turns into the retarded function
\begin{equation}
[G^R (\epsilon)]_{P_b(U)}=\frac{1}{U_2-U_1}
ln \left[
\frac{2\epsilon-U_2}{2\epsilon-U_2}\frac{2\epsilon-U_1}{2\epsilon+U_1}
\right]
\end{equation}
Depending on the parameters of the box distribution the logarithmic cuts 
may either overlap or remain separated. In the latter case the Mott
transition induced by sufficiently strong fermion hopping
will occur; the case of a local limit with arbitrarily small gap
should help to study the Mott transition also arbitrarily close to the 
local limit.

\subsubsection{Some physical quantities related to the gapped fermion 
density of states for $T<T_f$}
Below the freezing temperature $O(q^2)$--corrections  occur in $<\rho>$.
Using the exact low temperature solutions of section 2.2 one can obtain 
some information about the density of states deep in the ordered phase.
From the relation $[\nu]_{av} = 2 - \tilde{q}$
(valid in this form only at $ T = 0$)
and $\tilde{q}(T = 0, 0 < \mu < \frac{1}{\sqrt{2 \pi}})=0$ follows that the
system stays half filled independent of the chemical potential. 
This quantity can also be obtained from the disorder averaged 
one particle density of states by
\begin{equation}
[\nu]_{av}(\mu) = \sum_{\sigma = \uparrow, \downarrow} \int_{-\infty}
^{\infty}
 d\epsilon f(\epsilon) [\rho]_{av} (\epsilon, \mu)
\end{equation}
and hence display consequences of the interaction induced gap 
around zero energy.\\
The situation may be comparable to the antiferromagnetic gap in the clean
Hubbard model \cite{rozenberg}.

\pagebreak


\section{QUANTUM FIELD THEORY WITH BROKEN REPLICA PERMUTATION SYMMETRY}
This chapter describes the central results of the present paper. 
The Parisi symmetry, 
described in \cite{rammal} in terms of symmetric groups and more commonly
known for its ultrametric structure, is now shown to characterize a 
quantum field theory. The problems which arise in the QFT due to the 
effect of
RPSB are first resolved in a one--step breaking scheme. Then general
relations for arbitrary steps of RPSB are found in addition, which allow
conclusions on the low temperature behaviour.\\
The fermion Green's function can be derived, like any other correlation 
function and at any order K of Parisi--RPSB, from the generating functional
$\Xi$. The one--fermion Green's function can be obtained 
according to
\begin{equation}
{\cal{G}}_{i,j,\sigma}^{aa}(\epsilon_l)=
\frac{\delta}{\delta\eta^{a,l}_{i,\sigma}}
\frac{\delta}{\delta\bar{\eta}^{a,l}_{j,\sigma}}ln
\Xi_n(\{\eta\},\{\bar{\eta}\})
\end{equation}
employing derivatives with respect to the generating fields ${\eta}$ and 
$\bar{\eta}$.
This functional involves gaussian averages over (K+1) fluctuation fields 
$z_{\gamma}$. Since details are given in Appendix 1, we simply introduce 
its structure by the following symbolic shorthand notation
\begin{equation}
\Xi_n(\eta,\bar{\eta})=e^{-\frac{N}{4}\beta^2 J^2 Tr
Q_{Parisi}^2}\left[\prod\int^G_{z_{\gamma}}\right]
\int{\cal{D}}\Psi
Exp\left[
\bar{\Psi} g^{-1}\Psi+\eta\bar{\Psi}-
\bar{\eta}\Psi\right]
\end{equation}
where
\begin{equation}
g_{\sigma}(\underline{p},\epsilon_l|\{z_{\gamma}\} )=
(i\epsilon_l+\mu+\sigma\tilde{H}(\{z_{\gamma}^{\alpha_{\gamma}}\}))^{-1} 
\end{equation}
denotes the unaveraged fermion Greens function of the localized 
fermion exposed to the statistically fluctuating effective field \\
\begin{equation}
\tilde{H}(\{z_{\gamma}^{\alpha}\})= H + \sum_{\gamma}
\sqrt{q_{\gamma}-q_{\gamma+1}}\hspace{.1cm} z_{\gamma}^{(\alpha_{\gamma})}
\end{equation}
which depends on all fluctuation fields $z_{\gamma}$. The fermion--propagator 
of the insulating models is of course local in real space.
Spin (decoupling)--fields $z_{\gamma}$, carrying a Parisi block index
$\gamma$ ($\alpha_{\gamma}$ runs over all replicas which belong to a 
particular Parisi block), explore the random magnetic order. 
In the metallic case, the bare Green's 
function $g_0=g(\tilde{H}=0)$ and hence g itself become nonlocal, a 
complication that need not be studied now in order to understand the Parisi
symmetry of the QFT.\\
The Parisi matrix $Q_{Parisi}$ has the wellknown form 
\cite{parisi1} apart from the nonvanishing diagonal elements
$\tilde{q}$ (which we identify with $q_0$);   
their presence is required by the fact that $({\hat{\sigma}}^z)^2=
(\hat{n}_{\uparrow}-\hat{n}_{\downarrow})^2\neq 1$.   
The structure of the Parisi--matrix is of course responsible for
the rather complicated form of the Lagrangian (see Appendix 1); despite 
this
complication the fermion fields can be eliminated in the standard way,
which leads to the selfconsistent equations given below. 
\subsection{One--step replica permutation symmetry breaking}
It is instructive to start with a one step replica symmetry breaking 
(denoted by $K=1$), which is also considered to be the correct result in 
the regime of discontinuous thermal transitions. 
Here we wish to develop the full Parisi solution from this
first nontrivial starting point. Denoting by $q_1$ and by $q_2$ the
new offdiagonal block elements, the decoupling field z splits up into $z_1$
and $z_2$. 
The Green's function assumes the suggestive form
\begin{equation}
{\cal G}_{\sigma}(\epsilon_l)=
[[[(i\epsilon_l+\mu+\sigma\tilde{H}(z_1,z_2,y^a)^{-1})]_{y^a}]_{z_1}]_{z_2}
\end{equation}
which includes three successive averages to be specified below.
The instability of the replica--symmetric solutions becomes strongest as 
$T\rightarrow 0$, but only recently this has found its explicit 
representation in a zero temperature quantum field theory
including the low temperature effects. 
While quantum magnetic transitions are so far unaffected by
this instability to a large extent, the fermionic density of states and hence 
the fermion Green's function were observed to depend strongly on RPSB. 
This leads to a quantum field theory with Parisi symmetry,
ie with ultrametric structure. The fermion propagator had
acquired a nonsimple form already in the replica--symmetric approximation 
due to the random interaction on one hand and furthermore through
the effects of spin glass order. We now wish to show in more detail how this 
solution is improved at first--step breaking ($K=1$) and furthermore how a 
K--invariant relation can be derived to prove the one--fermion pseudogap 
reported in \cite{epl}.\\
We have recently presented the first solution to the question how Parisi 
replica permutation symmetry breaking (RPSB) and the related nonconstant part 
of the Parisi spin glass order parameter function $q(x)$ 
\cite{parisi1,BiYo,FiHe} are displayed in the low temperature many body 
theory of fermionic systems with frustrated Ising--interactions, 
emphasizing the $T=0$--limit in particular. 
Despite the fact that the interval $0\leq x\leq x_1$ of 
nonconstant $q(x)$ representing RPSB vanishes with temperature T as 
$T\rightarrow 0$, we find a large $O(T^0)$--effect to persist in many 
important physical quantities. This includes replica--diagonal fermion 
Green's function and fermion density of states, where at any step K of RPSB 
the set of different order parameters is seen to determine the
quantum--dynamical behaviour of the fermion propagator and of vertex 
functions. These effects are complementary to and not in contradiction with
recent replica--symmetric descriptions of $T=0$ quantum spin glass transitions
\cite{sro}. Parisi--RPSB \cite{parisi1} is seen to decide the qualitative 
and quantitative features of the low energy excitation spectrum.
While results are presented for an insulating model, the effect appears to be
rather model--independent and should hence be felt in transport 
properties of models with additional hopping hamiltonian for example.\\ 
We wish to provide results which evidence the fact 
that fermionic spin glasses also link closely glassy magnetic order and 
transport behaviour; further similarities between Hubbard model and the 
fermionic spin glass have been traced back to the particular role of the 
Onsager--Brout--Thomas reaction field \cite{BiYo,vollhardt}
for all these systems, as can be observed by comparing Hubbard--CPA-- 
\cite{vollhardt} with fermionic TAP--equations \cite{robr}.\\
Spin-- and charge--excitation spectra of fermionic spin glasses must be 
evaluated in order to construct a meaningful many body theory.
This article focusses on the effect of Parisi replica permutation symmetry 
breaking (RPSB) on the single fermion density of states (DoS), 
hence on the fermionic Green's function, and, by virtue of the Ward
identity for charge conservation, also on vertex functions, thus on the entire
ensemble of quantities that provide the basis of many body theories for 
fermionic systems with frustrated interactions.\\
It is known that replica--diagonal quantities like the linear equilibrium 
susceptibility $\chi$ feel Parisi symmetry breaking even at $T=0$ 
\cite{parisi2,BiYo,FiHe} despite 
the fact that the nontrivial part of the Parisi function only lives on an 
interval of width T.\\ 
The susceptibility had been analysed by Parisi for the standard SK--model 
who found a rapid convergence towards the exact result as the number of 
order parameters increased, this number being equal to $K+1$ in the SK--
and equal to $K+2$ in fermionic models. 
While the low temperature regime of the SK--model had not been of particular 
interest from the point of view of phase transition theory, it becomes highly
important for fermionic spin glasses, since the $T=0$--theory of excitation 
spectra plays a crucial role and, for the additional reason that some 
models exhibit quantum phase transitions along the $T=0$--axis. 
Parisi nevertheless analysed the low T regime \cite{parisi2} of the 
SK--model finding that K--step RPSB on one hand 
provided increasingly good approximations but failed to completely remove 
the negative entropy and the instability problem at low enough 
temperatures unless $K\rightarrow \infty$.\\ 
In this paper the effect of one step RPSB on the density of states is 
presented in detail, followed by the derivation of an analytical relation 
valid for all K, 
which allows to determine the type of excitation spectrum present in the 
full Parisi solution for the fermionic Ising spin glass.
Despite the fact that the regime of deviation from a replica--symmetric
spin glass order parameter is only of $O(T)$, we find that it has a large 
$O(T^0)$--effect on the density of states, the one--particle--,
and  many--particle Greens functions at $T=0$. This density of states is 
derived as usual from the imaginary--time (disorder--averaged) fermion 
Green's function $[<{\bar{\psi}}(\tau)\psi(\tau^{\prime})>]_{av}$, which is
one of the decisive quantum--dynamical elements of any many body 
theory of fermionic spin glasses. This illustrates that, unlike the usual
picture of a Parisi solution being just a static order parameter function,
the fermionic picture must include the qualitative extension to 
dynamical quantities. Those become drastically altered by the nontrivial 
part of the Parisi solution, which is otherwise invisible at $T=0$, hence 
providing a {\it quantum--dynamical image of RPSB}.\\
Let us consider the Parisi solution of the infinite--range fermionic Ising 
spin glass defined as the SK--model extension in a grand canonical 
ensemble with spin one half operators represented as 
$\hat{\sigma}^z=\hat{n}_{\uparrow}-\hat{n}_{\downarrow}$. \\
\subsubsection{Fermionic density of states}
It is known since Parisi's work \cite{parisi2} that an analytical 
low temperature expansion is hard to obtain even for the standard SK--model 
and its smaller set of selfconsistent parameters.
First insight is gained by the one--step RPSB ($K=1$).
The standard three parameter set of the SK--model for $K=1$, 
order parameters $q_1$ and $q_2$, and $m\equiv m_1\sim T$, 
is enlarged in the fermionic space by $\tilde{q}-q_1\sim T$, where
$\tilde{q}:=[<\sigma(\tau)\sigma(\tau^{\prime})>]_{av}$ represents
a spin correlation, which remains static unless a fermion hopping 
mechanism or other noncommuting parts are included in the Hamiltonian.
For the fermionic Ising spin glass the ($K=1$)--DoS reads 
\begin{equation}
\rho_{\sigma}(E,H)=\frac{ch(\beta\mu)+ch(\beta 
E)}{\sqrt{2\pi(\tilde{q}-q_1)}}\frac{e^{-\frac{1}{2}\beta^2 
J^2(\tilde{q}-q_1)}}{\sqrt{2\pi q_2}}
\int_{-\infty}^{\infty}dv_2 
e^{-\frac{v_2^2}{2 q_2}}\frac{\int_{-\infty}^{\infty}dv_1
e^{-\frac{(v_1-v_2)^2}{2(q_1-q_2)}-\frac{(v_1+H+\sigma 
E)^2}{2(\tilde{q}-q_1)}}{\cal{C}}^{m-1}}{\int_{-\infty}^{\infty}dv_1 
e^{-\frac{(v_1-v_2)^2}{2(q_1-q_2)}}{\cal{C}}^m}
\label{65}
\end{equation}
where
\begin{equation}
{\cal{C}}=cosh(\beta \tilde{H})+\zeta,\quad\quad
\zeta=cosh(\beta\mu)exp(-\frac{1}{2}\beta^2(\tilde{q}-q_1)).
\label{defC}
\end{equation}
This expression reveals the competition between the particle "pressure" 
exerted by the chemical potential $\mu$ and the single--valley susceptibility 
\begin{equation}
\bar{\chi}=\beta(\tilde{q}-q_{_1})\nonumber
\end{equation}
leading to a crossover 
at $|\mu|=\frac{1}{2}\bar{\chi}$ in the $T\rightarrow 0$--limit. 
The $\zeta$--term is a fermionic feature which is absent from the standard 
SK--model. It is closely related to the fermion filling; this filling factor 
behaves discontinuously on the $T=0$--axis \cite{brro}.
The zero temperature limit allows to perform the $v_1$--integrations, 
which results in the exact $T=0$ formula for the density of states
\begin{equation}
\rho_{\sigma}(E)=
\frac{e^{-\frac{1}{2}a^2(H)(1-q_2)-\frac{\Delta_E^2}{1-q_2}+a(H)\Delta_E-
\frac{H^2}{2 q_2}}}{\pi\sqrt{1-q_2(H)}}
\int_{-\infty}^{\infty}dz\frac{e^{-\frac{1}{2}\frac{z^2}{1-q_2}-
(\frac{\sqrt{q_2}}{1-q_2}\frac{\sigma E}{|E|}\Delta_E
-\frac{H}{\sqrt{q_2}})z}}{d(z)+d(-z)}
\Theta(\Delta_E)
\label{4}
\end{equation}
using the definitions
\begin{eqnarray}
d(z)&\equiv& e^{a(H)\sqrt{q_2}z}\left[1+Erf[\frac{a(H)(1-q_2)+
\sqrt{q_2}z}{\sqrt{2(1-q_2)}}]\right],\\
\Delta_E&\equiv& |E|-w(H),\\
a(H)&\equiv&lim_{_{T\rightarrow0}} \frac{d}{dT}m(T,H).
\end{eqnarray}
These equations allow to derive the first improved approximation beyond the 
replica--symmetric solution, given in the preceding section, which 
displays a magnetic hardgap of width $2 E_g(H)$ in the DoS and the fact that 
half--filling at $T=0$ extends over the finite interval of chemical 
potentials given by $|\mu|<\frac{1}{2}E_g(H)$. Beyond this interval phase 
separation occurs together with a discontinuous transition into a full or 
an empty system \cite{brro}. More details about the phase diagram are 
presented in the subsequent paper II.\\ 
A stable and spatially--homogeneous saddle--point solution 
could only be found for the half--filled case. Thus the following analysis is 
restricted to this interval of chemical potentials. Its width is determined 
selfconsistently and seen to decrease to zero as K tends to infinity.
For $|\mu|\leq\frac{1}{2}\lim_{T\rightarrow 0}(\beta(\tilde{q}-q_1))$ we first 
derive the coupled set of selfconsistent equations
\begin{eqnarray}
\tilde{q}&=&1-\int_2\frac{\int_1 {\cal{C}}^{m-1} \zeta}{\int_1 
{\cal{C}}^m}\\
q_1&=&\int_2\frac{\int_1 {\cal{C}}^{m-2} sinh^2(\beta \tilde{H})}{\int_1 
{\cal{C}}^m}\\
q_2&=&\int_2 \left[\frac{\int_1 
{\cal{C}}^{m-1}sinh(\beta\tilde{H})}{\int_1 
{\cal{C}}^m}\right]^2\\
0&=&\frac{\partial}{\partial m} f=\frac{\beta J^2}{4}(q_1^2-q_2^2)+
\frac{T}{m^2}\int_2 ln\int_1{\cal{C}}^m - 
\frac{T}{m}\int_2\frac{\int_1{\cal{C}}^m ln{\cal{C}}}{\int_1{\cal{C}}^m} 
\end{eqnarray} 
The shorthand gaussian integral notation 
\begin{equation}
\int_l^G\equiv\int_{-\infty}^{\infty}\frac{du_k}{\sqrt{2\pi(q_k-q_{k+1})}}
exp\left[-\frac{u_k^2}{2(q_k-q_{k+1})}\right]
\label{defint_k}
\end{equation} 
adopts the normalization used by Parisi \cite{parisi1}.
For $T=0$, we solve the selfconsistency equations up to one final 
integral, finding the $T=0$--set of coupled equations
\begin{eqnarray}
& &\hspace{-.5cm}\tilde{q}=q_1=1,\hspace{.2cm}
lim_{T\rightarrow0}\frac{\tilde{q}-q_1}{T}=\bar{\chi},\hspace{.2cm}
q_2=\int_{-\infty}^{\infty}\frac{dz}{\sqrt{2\pi}}
e^{-\frac{(z-H/\sqrt{q_{_2}})^2}{2}}\left[\frac{d(z)-
d(-z)}{d(z)+d(-z)}\right]^2
\\
& &\hspace{-.5cm}0=1-q_2^2-\frac{4}{a}
\int_{-\infty}^{\infty}\frac{dz}{\sqrt{2\pi
}}\hspace{.1cm}e^{-\frac{(H/\sqrt{q_{_2}}-z)^2}{2}}
\{-\frac{1}{a}\hspace{.1cm}ln[\frac{1}{2}e^{\frac{1}{2}a^2 t}
(d(z)+d(-z))]\nonumber\\
& &\hspace{-.3cm}+\hspace{.1cm}
\left[(at+\sqrt{q_{_2}}\hspace{.1cm}z)d(z)+
(at-\sqrt{q_{_2}}\hspace{.1cm}z)d(-z)+\sqrt{8
t/\pi}\hspace{.1cm}e^{-\frac{1}{2}a^2
t-\frac{1}{2}q_{_2} z^2/t}\right]/[d(z)+d(-z)]\}\\
\nonumber
\end{eqnarray}
where $t\equiv q_1-q_2$.
All parameters represent the temperature-- and magnetic 
field--dependent solutions. \\ 
The solutions for $T=0$ and $H=0$ are given by
\begin{eqnarray}
q_2&=&0.476875\\ 
a&=&lim_{T\rightarrow 0}\frac{d}{dT}m(T)=1.36104\\
w&=&lim_{T\rightarrow 0}\frac{\tilde{q}-q_1}{T}=.239449.
\end{eqnarray}
\begin{figure}
\epsfig{file=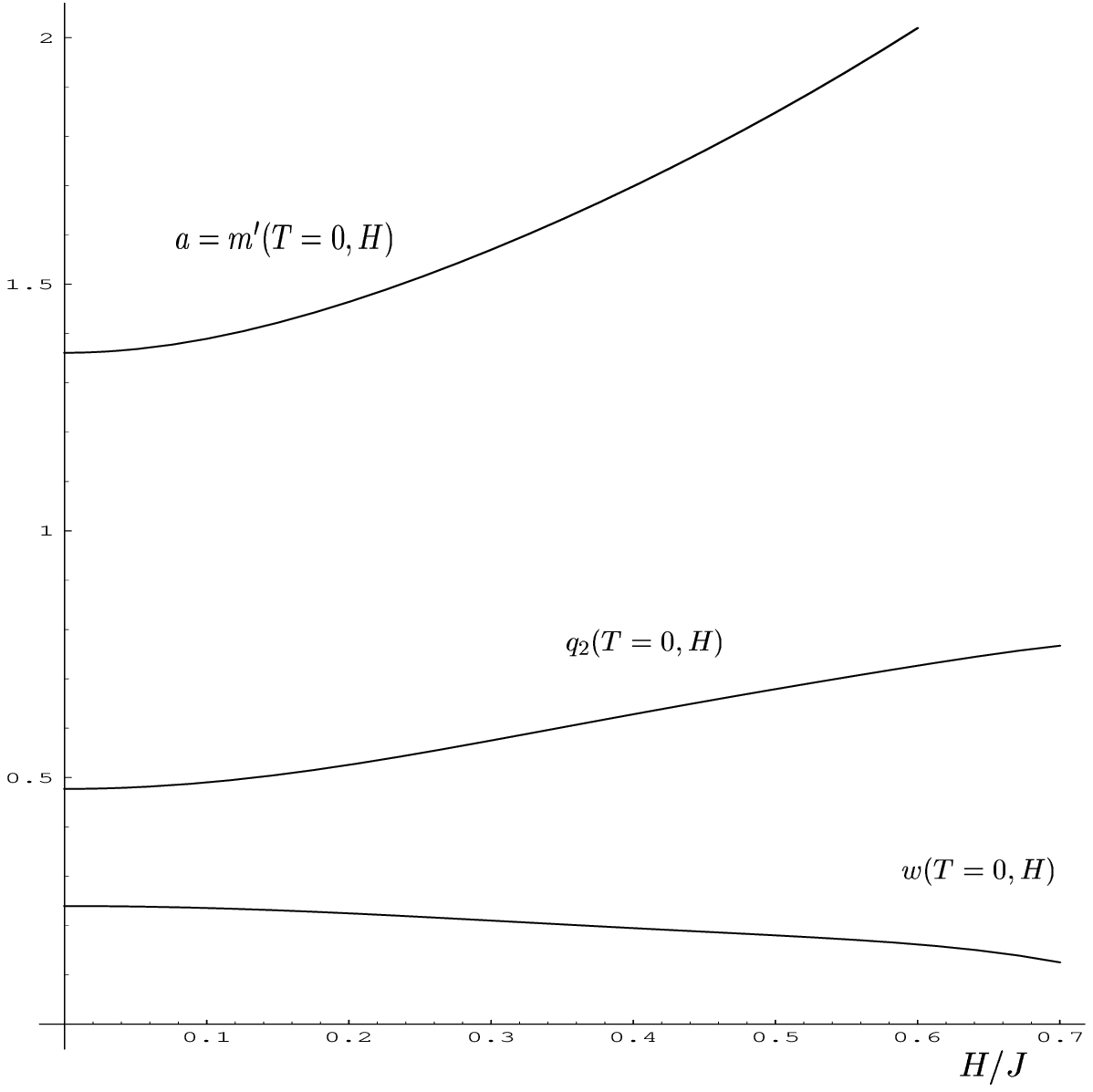,width=7.7cm,angle=270}
\caption{Field dependence of $dm/dT$ (top),
of the order parameter $q_2$, and of gapwidth parameter w (bottom) for 1RPSB
and zero temperature.}
\hspace{2cm}
\epsfig{file=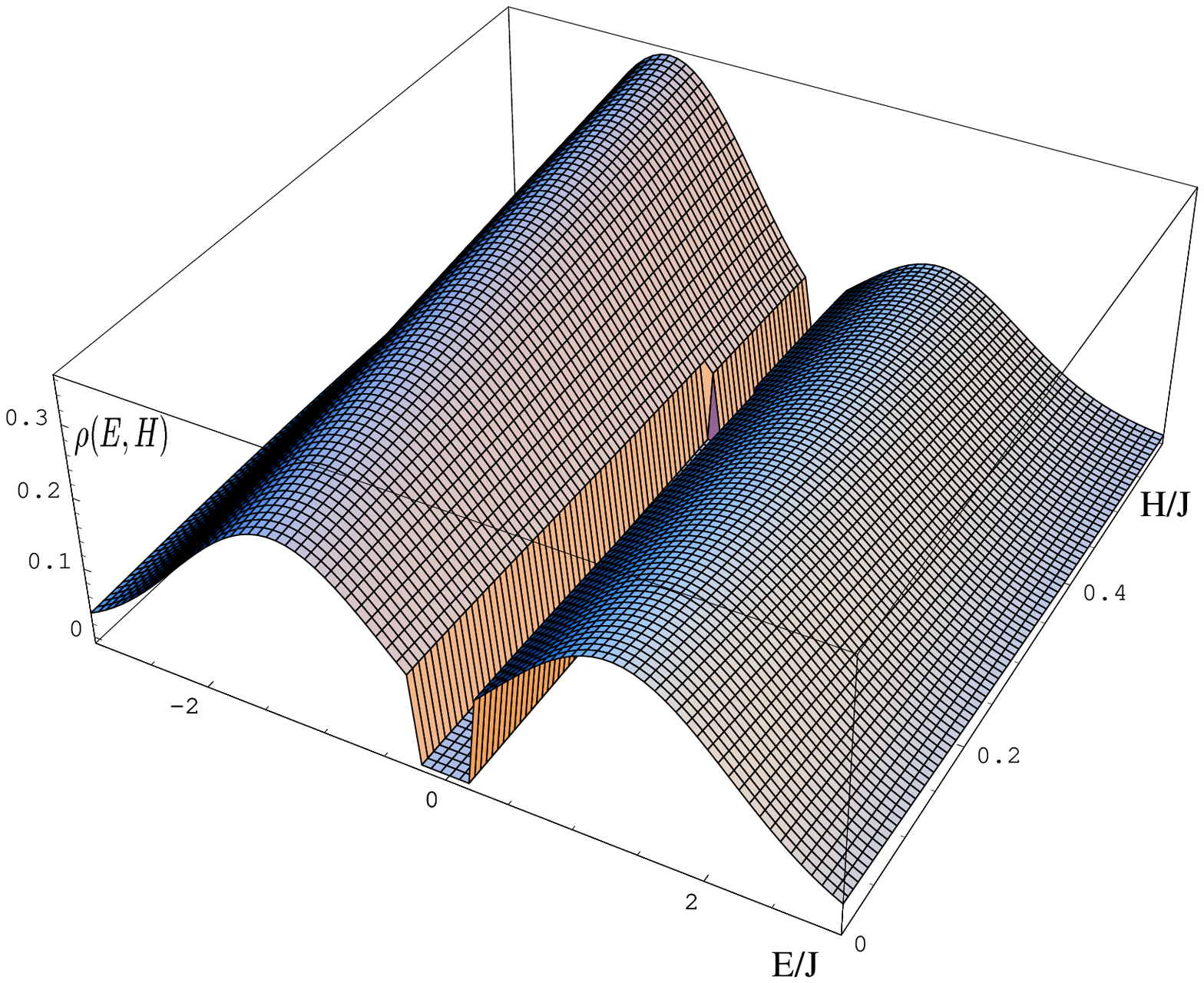,width=7.7cm,angle=270}
\caption{Zero temperature density of states as a function of energy and
magnetic field for 1--step RPSB}
\end{figure}
The H--dependent 
solutions shown in fig.6 are then used in evaluating eq.(\ref{4}) 
for the density of states. $T=0$--results are shown in figs.7 and 8, while
the result at finite low temperature of fig.9 illustrates the presence of 
plateaus of constant slope, each corresponding to Parisi order parameter
separations (here: $\tilde{q}-q_1$ and $q_1-q_2$). The number of these 
plateaus of constant slope increases with the order K of Parisi--RPSB.
Hence, the time--dependence of the Green's function should 
characteristically depend on the order parameter separations 
${q_k-q_{k-1}}$.
\par
\begin{figure}
\hspace{2cm}
\epsfig{file=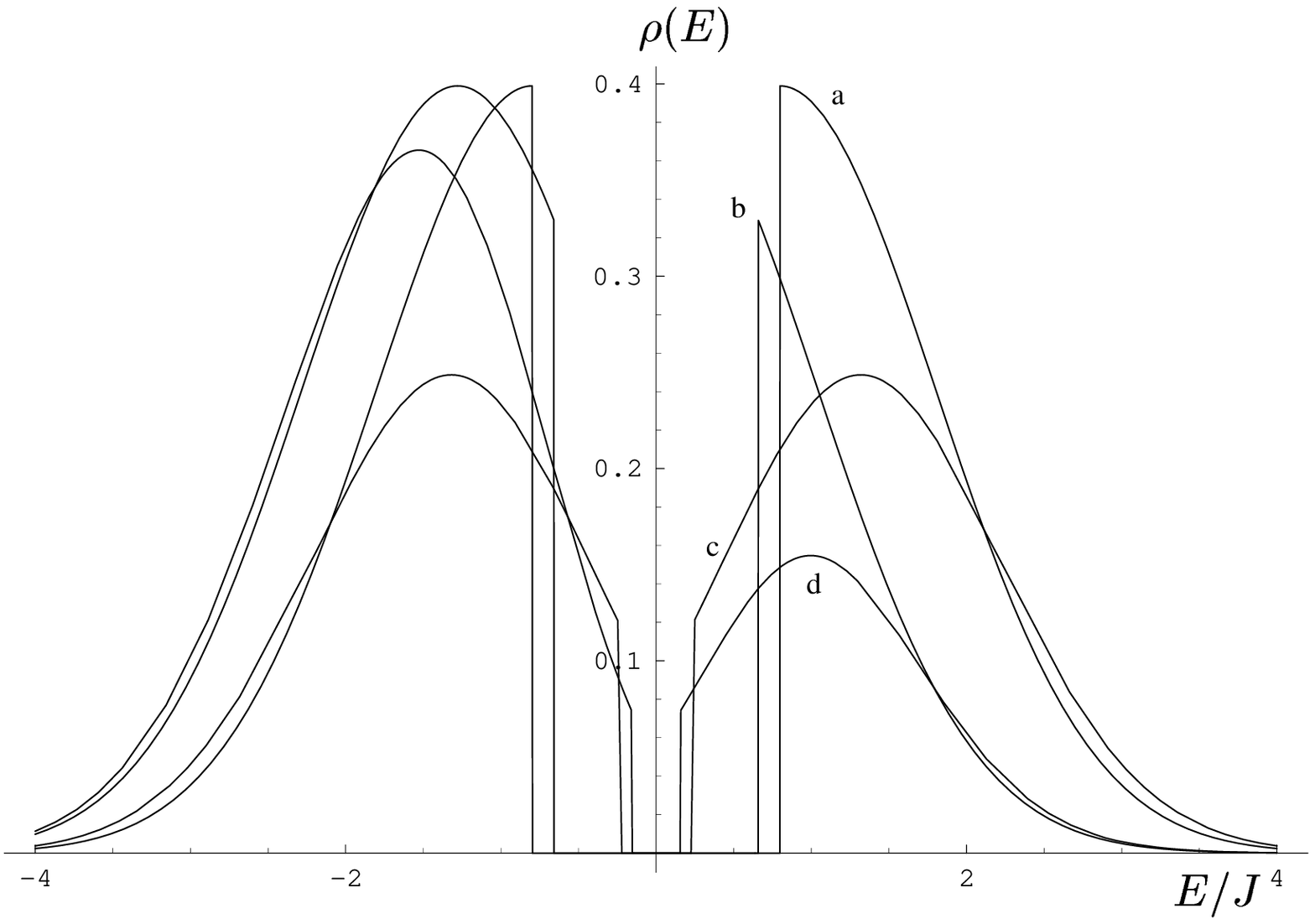,width=7cm,angle=270}
\caption{Effect of one step replica symmetry breaking on the fermionic DoS for
magnetic fields $H=0$ (curve c: 1RPSB, a: 0RPSB) 
and $H/J=0.6$ (d: 1RPSB, b: 0RPSB)}
\hspace{2cm}
\epsfig{file=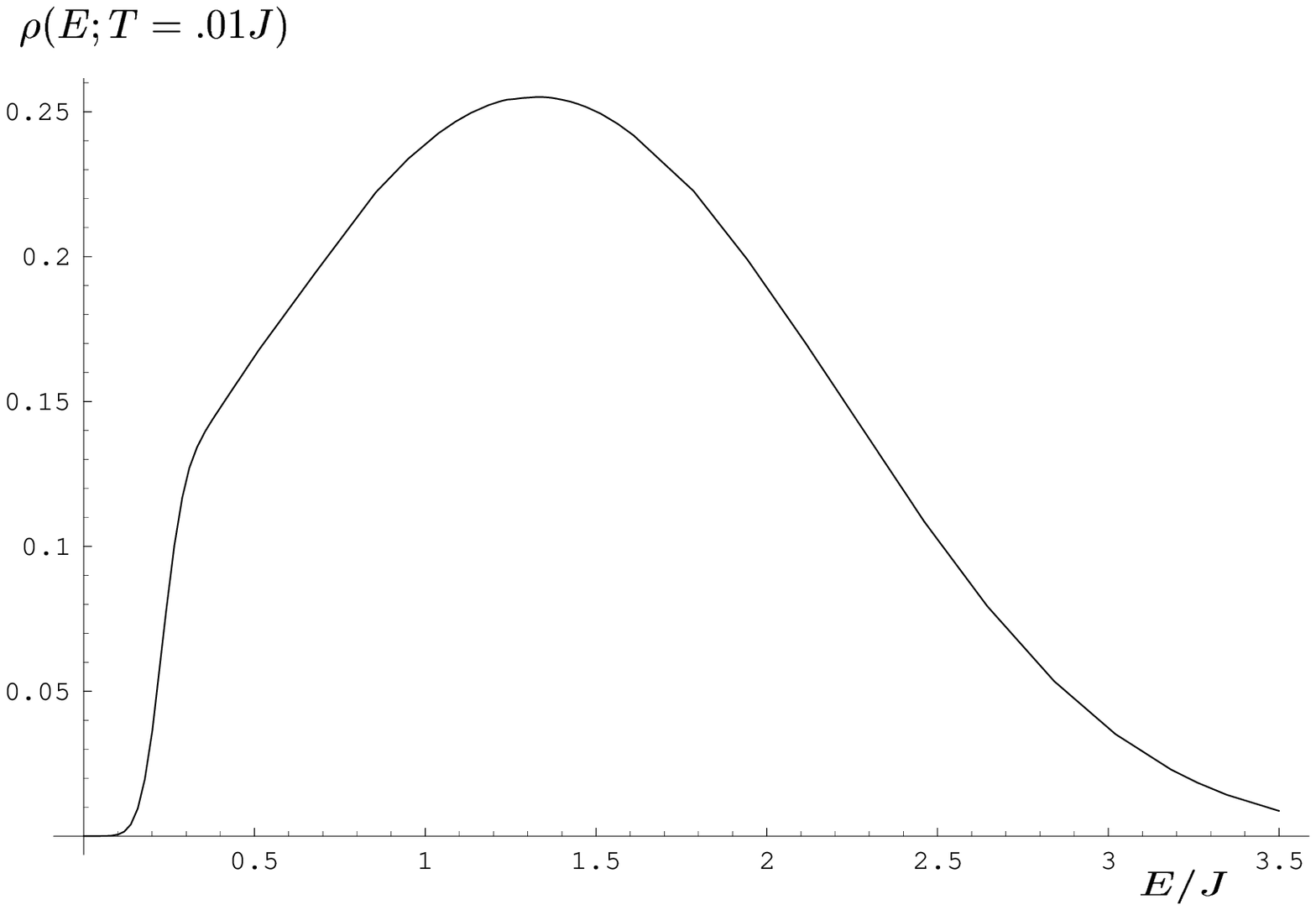,width=7cm,angle=270}
\caption{The 1--step RPSB solution for the zero field density of states at 
very low temperatures ($T=.01 J$) ($\rho$ is symmetric w.r.t. E). The 
$T=0$ hardgap of reduced size is already digged out, yet two different 
slopes $\frac{d\rho}{dE}$ corresponding to $q_0-q_1=O(T)$ (steep portion) 
and to $q_1-q_2=O(1)$ remain visible} 
\end{figure} 
\begin{figure}
\hspace{2cm}
\epsfig{file=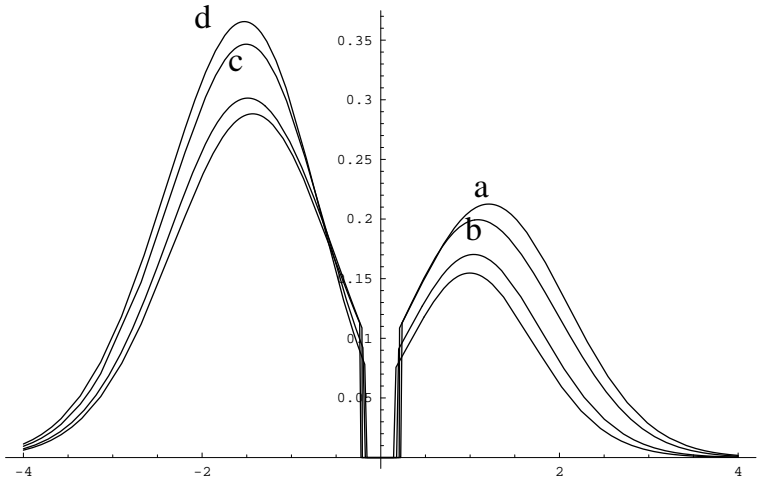,width=7cm,angle=270}
\caption{DoS at 1--step RPSB for magnetic fields $H/J=.2$ (curve a), $.3$ 
(b), $.5$ (c), and $.6$ (d)}
\end{figure}
If we compare with the replica--symmetric result a reduction of 
the gapwidth is observed. Analytically one finds a gapwidth
\begin{equation}
E_g (H)=\bar{\chi}=lim_{T\rightarrow 0}\beta(\tilde{q}-q_{_1})
\end{equation}
which turns into $lim_{T\rightarrow 0}(\beta(\tilde{q}-q(1)))$ 
in terms of the Parisi function $q(x)$ at $K=\infty$. 
Only in absence of RPSB this susceptibility coincides with the equilibrium 
one, denoted by $\chi$. 
In fact for 1--step RPSB the fermionic Ising spin glass approaches 
$\chi=\beta(\tilde{q}-q_1)+\beta m(q_1-q_2)\rightarrow .95$, 
the same numerical value as the one for the SK--model.\\
\subsubsection{The fermion propagator}
We are interested in the explicit form of the fermion propagator at 
one--step RPSB for several reasons. To mention a few, this function 
${\cal{G}}^{-1}(\epsilon_l)$ 
is an indispensable element of the quantum field theory with Parisi symmetry,
it is directly linked to some observables, it contributes by Re(G) a second
quantity which illuminates the crossover to a pseudogap, it is also 
needed to formulate the Ward identity, and it adds another
challenge for the mapping on a random U Hubbard model in the local limit.
We start out from Eq.(\ref{greensfunction}) (as derived in Appendix 1) 
which yields in the 
replica limit for 
$K=1$ 
\begin{equation}
{\cal{G}}^{(1)}_{\sigma}(\epsilon_l)=\int_2\frac{\int_1 
{\cal{C}}^{m-1}\int_0 
e^{\Phi}\frac{1}{i\epsilon_l+\mu+\sigma\tilde{H}(z_0\equiv 
y,z_1,z_2)}}{\int_1 {\cal{C}}^m},
\end{equation}
where $exp(\Phi)=2(cosh(\beta\tilde{H})+cosh(\beta\mu)e^{-\frac{1}{2}\beta
\overline{\chi}})$.\\
In the regime $\mu<\frac{1}{2}\overline{\chi}$ one obtains 
\begin{eqnarray}
{\cal{G}}^{(1)}(\epsilon_{_l})&=& -i\sqrt{\frac{\pi}{2}} \int_{z_{_2}}^G 
\frac{\frac{1}{\tilde{q}-q_{_1}}
}{\int_{z_1}^G{\cal{C}}^m}\int_{z_{_1}}^G {\cal{C}}^{m-1}
[2 
cosh(\beta\mu)e^{\frac{(\epsilon_{_l}-i(\mu+\tilde{H}_{_0}))^2}{2(\tilde{q}
-q_{_1})}}\left[1-Erf(\frac{\epsilon_{_l}
-i(\mu+\tilde{H}_{_0})}{\sqrt{2(\tilde{q}-q_{_1})}})
\right]\nonumber\\
&+&\sum_{\lambda=\pm1}e^{\lambda\beta\tilde{H}_{_0}+
\frac{1}{2}\beta\overline{\chi}}e^{\frac{(\epsilon_l-i(\mu+\tilde{H}_{_0}
\overline{\chi}))^2}{2(\tilde{q}-q_{_1})}}
\left[1-Erf(\frac{\epsilon_{_l}
-i(\mu+\tilde{H}_{_0}
+\lambda\overline{\chi})}{\sqrt{2(\tilde{q}-q_{_1})}})\right]]
\end{eqnarray}
The zero temperature limit (using $lim_{T\rightarrow 0}\beta m=a$ finite)
simplifies this result and can be cast into the form
\begin{equation}
{\cal{G}}^{(1)}_{\sigma}(\epsilon)=\int_{-\infty}^{\infty}
\frac{dz_{_2}}{\sqrt{2\pi}} 
exp(-\frac{1}{2}z_2^2)
\frac{\int_{-\infty}^{\infty}dz_1 
\Theta(\tilde{H}_0)e^{-\frac{1}{2}z_1^2} 
e^{a\tilde{H}_0}\left[\frac{1}{i\epsilon+\mu+\tilde{H}_0+\overline{\chi}}+
\frac{1}{i\epsilon+\mu-\tilde{H}_0-\overline{\chi}}\right]}{
\int_{-\infty}^{\infty}dz_1 
e^{-\frac{1}{2}z_1^2+a|\tilde{H}_0|}}
\label{72}
\end{equation}
where $\tilde{H}_0\equiv\tilde{H}(0,z_1,z_2)$. The spectral 
representation (\ref{spectral}) now employs the 1--step RPSB--result for 
the density of states Eq.(\ref{65},\ref{4}).\\
Before evaluating the $z_1$--integrals we can reconsider the mapping 
with the local limit of a random U Hubbard model and the changes that 
occur due to RPSB in the spin glass. 
Now 
\begin{eqnarray}
<U>&\rightarrow& 2\hspace{.2cm}\overline{\chi}\\
\frac{\delta U}{2}&\rightarrow& 
\tilde{H}_{_0}=\sqrt{q_1-q_2}\hspace{.2cm}z_{_1}+\sqrt{q_2}\hspace{.2cm}z_{_2} 
\end{eqnarray}
The second correspondence shows that only positive $U$-fluctuations around 
a positive mean value occur, which decreases with each step of RPSB like the 
gapwidth of the fermionic spin glass. However, modelling the latter by 
the random $U$ interaction requires ($K+1$) fluctuation--fields $\delta 
u_{\alpha}$ at $K$--th order RPSB and subject to the constraint 
$\delta U\equiv\sum_{\alpha}\delta u_{\alpha}>0$. Apart from the overall 
condition of positive $U$ and $\delta U$ the fluctuation--fields are 
allowed to assume any negative value. Let us express the speculation that
the existence of realizations of arbitrarily negative attractive 
interactions might open the way to a pairing creation due to Parisi 
symmetry breaking. These pairs might become delocalized due to the 
introduction of a fermion hopping term. We do not pretend to have an answer
at the moment, but we think it is justified to raise the question {\it 
whether a superconducting transition might become possible due to the Parisi 
replica symmetry breaking} at and beyond the instability line towards RPSB.\\
The random $U$ local Hubbard limit was constructed to match the fermionic 
spin glass propagator at $T=0$. It is clear that this mapping cannot be 
achieved selfconsistently for all temperatures, since the zero dimensional 
local limit does not allow for a thermal phase transition. We consider it 
possibly important that one may achieve a mapping at zero temperature between
a random $U$ Hubbard model and a metallic Ising spin glass with identical
hopping hamiltonians. This will involve the question of replica symmetry
breaking in the random $U$ Hubbard model. \\
The role of frustration and the analog of the Almeida Thouless instability
in the random $U$ Hubbard model are to be studied, which requires to 
study the replicated field theory with incomplete or nongaussian disorder
average. This is a problem for later studies.\\ We also consider 
interesting the question, whether frustration in clean systems leads to
time--dependent behaviour that resembles the one generated by replica
symmetry breaking. 
\subsubsection{Quantum--dynamical image of RPSB}
\begin{figure}
\hspace{2cm}
\epsfig{file=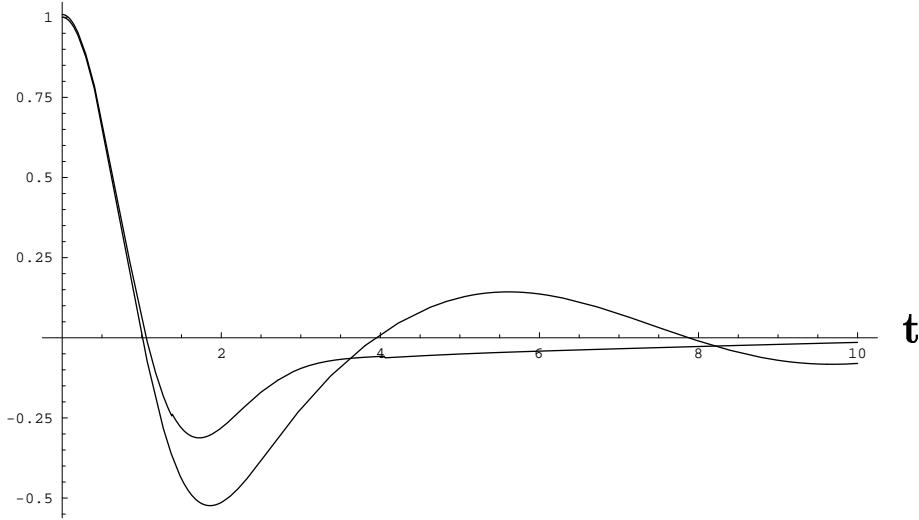,width=8cm,angle=270}
\caption{Comparison of short--time dependence of the replica--symmetric 
retarded Green's function $i e^{-1\mu t}G^{(0),R}(t)$ and of the 1--step 
broken $i e^{-i\mu t}G^{(1),R}(t)$} 
\label{f11}
\end{figure}

\begin{figure}
\hspace{2cm}
\epsfig{file=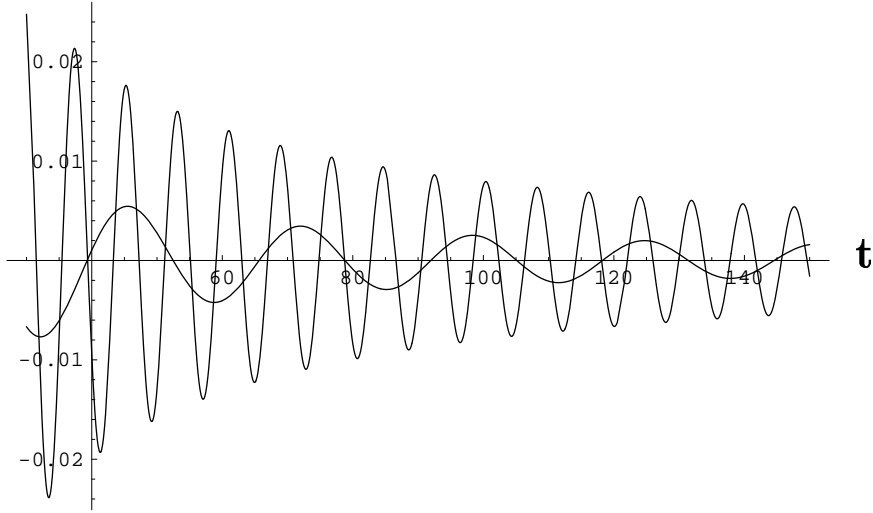,width=8cm,angle=270}
\caption{Long--time decay of the fermion Green's function showing 
gap--induced faster (0RPSB) and slower (1RPSB) 
oscillations. The amplitude c of the $\frac{c}{t}$ long time decay 
is reduced by 1--step RPSB. Oscillations and amplitude c vanish as 
($K\rightarrow\infty$)--step RPSB (not shown).} 
\label{f12}
\end{figure}

The results for low energy excitations indicate that the 
time--dependence of fermion correlation functions must be affected by 
replica symmetry breaking. We study now the fermion propagator and in 
particular the retarded Green's fucntion. The Fourier transformation
\begin{equation}
G^R(t)=\int^{\infty}_{-\infty}\frac{d\epsilon}{2\pi} e^{-i\epsilon t} 
G^R(\epsilon)\nonumber 
\end{equation}
yields
\begin{equation}
G^{(1)R}(t)=-i e^{i\mu t}\{\sqrt{\frac{2}{\pi q_2}}e^{-\frac{1}{2}
a^2(q_1-q_2)} \int^{\infty}_{-\infty}du
\frac{e^{-\frac{1}{2}(\frac{1}{q_2}+
\frac{1}{q_1-q_2})u^2+{\cal{R}}^2(u,t)}
\left[1+Erf\left[{\cal{R}}(u,t)\right]
\right]e^{i\hspace{.1cm}\overline{\chi}\hspace{.1cm}t}}{e^{a 
u}\left[1+Erf\left[{\cal{R}}(u,0)\right]\right]+e^{-a 
u}\left[1+Erf\left[{\cal{R}}(-u,0)\right]\right]} 
+c.c.\}
\end{equation}
where
\begin{equation}
{\cal{R}}(u,t)\equiv\sqrt{\frac{q_1-q_2}{2}}(a+\frac{u}{q_1-q_2}+it).
\end{equation}
This result is to be compared with the replica--symmetric result 
evaluated as 
\begin{equation}
G^{(0)R}(t)=e^{-\frac{t^2}{2}}\left[cos(\chi_{_0} t)
-Erfi(\frac{t}{\sqrt{2}})sin(\chi_{_0} t)\right]
\frac{e^{i\mu t}}{i}\Theta(t) 
\label{75}
\end{equation}
Both results are compared with each other for short times in Fig.(\ref{f11})
and for long times in Fig.(\ref{f12}).\\
For large times we obtain
\begin{equation}
G^{(1),R}(t\rightarrow\infty)\sim\frac{c(q_1,q_2,a)}{t}
\end{equation}
with an amplitude smaller than the one of the replica--symmetric solution 
(\ref{75}), which gives 
\begin{equation}
G^{(0),R}(t\rightarrow\infty)\sim 
-i\sqrt{\frac{2}{\pi}}\frac{sin(\chi_{_0}t)}{t} 
\end{equation}
The behaviour at large times of this $T=0$--quantum theory is obviously 
marked by RPSB. As will become clear below the susceptibility 
$\overline{\chi}$, on which the $O(\frac{1}{t})$--term of the fermion 
propagator depends, becomes smaller and smaller with increasing steps K of
RPSB and finally vanishes in the presumed exact solution at $K=\infty$.
The oscillations similarly become slower and disappear at $K=\infty$.\\
One may in this context recall the correspondence between long--time 
behaviour in the
case of classical Glauber dynamics, which was seen to correspond to the 
nontrivial part of the Parisi function \cite{sompolinsky}.


\subsection{Higher order RPSB and $K=\infty$}
The result in 1--step RPSB, although still unstable towards higher RPSB,  
can be viewed as a much better approximation than 0--RPSB, since it
already contains features of the full parisi solution at $K=\infty$.
In this section we consider the RPSB equations at arbitrary K.
\subsubsection{K--invariant ratios} 
We now derive two K--invariant relations. The first one, between zero 
temperature gap and nonequilibrium susceptibility both calculated in K--th 
order RPSB, is contained in 
\begin{equation} E^{(K)}_g=\bar{\chi}^{(K)}=
lim_{T\rightarrow0}\beta(\tilde{q}^{(K)}-q^{(K)}_1). 
\end{equation}
In order to obtain the second invariant we compare the 
$T\rightarrow0$--limit of the selfconsistent equations for $\tilde{q}-q_1$
and for the DoS at arbitrary order K of RPSB.\\
The Green's function at K--th order RPSB can be written as
\begin{equation}
{\cal{G}}(\epsilon_l)=\int_{K+1}\zeta_{K+1}\int_{K}\zeta_{K}...
\int_2\zeta_2
\frac{\int_1{\cal{C}}^{m_1-1}\int_y 
e^{\Phi}g(\epsilon_l)}{\int_1{\cal{C}}^{m_1}} \label{ho1}
\end{equation}
where 
\begin{equation}
\zeta_{K+1}\equiv\left[\int_K\left[\int_{K-1}\left[\int_{K-2}...
(\int_1{\cal{C}}^{m_1})^{\frac{m_2}{m_1}}...\right]^{\frac{m_{K-1}}{m_{K-2}}}
\right]^{\frac{m_{_K}}{m_{_{K-1}}}}\right]^{-1}
\end{equation}
while $\int_K$ and ${\cal{C}}$ have been defined before by 
Eq.(\ref{defC}) and Eq.(\ref{defint_k}) respectively.\\ 
In the zero temperature limit Eq.(\ref{ho1}) turns into
\begin{equation}
{\cal{G}}(\epsilon)=\int_{K+1}\zeta_{K+1}|_{_{T=0}}\int_K\zeta_K|_{_{T=0}}
\hspace{.1cm}...\int_2
\frac{\int_1\Theta(\tilde{H_0})e^{a_1\tilde{H_0}}
[\frac{1}{i\epsilon+\mu+\tilde{H}_0+\bar{\chi}}+
\frac{1}{i\epsilon+\mu-\tilde{H}_0-\bar{\chi}}]}{\int_1e^{a_1|\tilde{H}_0|}}
\end{equation}
The density of states at the gap edge $E_g^{(K)}$ is hence given by
\begin{equation}
\rho^{(K)}(E_g^{(K)}+0)=\int_{K+1}\zeta_{K+1}|_{_{T=0}}\int_K\zeta_K|_{_{T=0}}
\hspace{.1cm}...\int_{-\infty}^{\infty}
\frac{du_{_2}e^{-\frac{(u_2-u_3)^2}{2(q_2-q_3)}}}{_{\sqrt{2\pi(q_2-q_3)}}} 
\frac{e^{-\frac{u_2^2}{2(q_1-q_2)}}}{\int_{-\infty}^{\infty}du_1 
e^{-\frac{(u_1-u_2)^2}{2(q_1-q_2)}+a_1|u_1|}} 
\end{equation}
A similar expression is obained for the $T=0$ gap energy $E_g^{(K)}$,
since
\begin{eqnarray}
E_g^{(K)}&=&lim_{_{T\rightarrow0}}\beta(\tilde{q}^{(K)}-q_1^{(K)})\nonumber\\
& 
&\hspace{-1.2cm}=\int_{K+1}\zeta_{K+1}|_{_{T=0}}\int_K\zeta_K|_{_{T=0}}\hspace{.2cm}...
\int_{-\infty}^{\infty}\frac{du_2
e^{-\frac{(u_2-u_3)^2}{2(q_2-q_3)}}}{_{\sqrt{2\pi(q_2-q_3)}}}
\frac{\lim_{_{T\rightarrow0}}\beta\int_{-\infty}^{\infty} du_1 
e^{-\frac{(u_1-u_2)^2}{2(q_1-q_2)}}cosh^{^{m_1-2}}(\beta 
u_1)}{\int_{-\infty}^{\infty} du_1 
e^{-\frac{(u_1-u_2)^2}{2(q_1-q_2)}+a_1|u_1|}} \end{eqnarray} 
Thus we obtain the second K--invariant ratio
\begin{equation}
\frac{lim_{_{|E|\downarrow E_g(H)}}\rho^{(K)}_{\sigma}(E)}{E^{(K)}_g}=
\frac{1}{2}
\label{10}
\end{equation}
Together with the fact that, for each given K, the gapwidth    
equals twice the fermionic nonequilibrium susceptibility $\bar{\chi}$, which
turns into $\bar{\chi}=\beta(\tilde{q}-q(1))$ differing only by
exponentially small terms from the spin--space--analog
$\bar{\chi}=\beta(1-q(1))\sim T$ \cite{BiYo},
where $q(1)$ denotes the Parisi function
$q(x)$ at $x=1$, the DoS--hardgaps at finite K terminate in a softgap   
for $K\rightarrow \infty$. Note that we did not have to evaluate the
$T=0$--Parisi function $q(x)$ in order to reach this conclusion.
Assuming that the relation between gapwidth and $\bar{\chi}$ remains valid
(at least in good approximation) for finite--range models, fluctuation 
effects should harden the gap. This requires further analysis.\\
While we have proved the existence of a spin--glass hardgap at any 
finite $K>0$ with
\begin{equation}
\delta\rho^{(K)}(T=0,E)\sim |E-E_g^{(K)}|
\end{equation}
for $|E|\geq E_g^{(K)}$ and $K$ finite, the pseudogap of the 
infinite--range model leads to
\begin{equation}
\rho(T=0,E)\sim |E|^x\quad (K=\infty)
\end{equation}
with a scaling exponent x, which could
eventually become different from one and remains to be determined.
This pseudogap together with x=1 would be slightly reminiscent of the 
exponent found for a superconducting glass unitary nonlinear sigma model 
\cite{ro}. However exponents differing from one are known even in mean 
fields solutions for gapless superconductors and also at our
spin glass thermal tricritical point \cite{brro}.
We remark that the pseudogap--solution given for the fermionic spin glass
model refers precisely to $\mu=0$ whereas the hardgap--solutions at
finite K happened to be stable within finite intervals $|\mu|\leq
\bar{\chi}/2$ corresponding to half--filling only at $T=0$.
The regime beyond half--filling, identified as the domain of phase
separation in the replica--symmetric solution \cite{brro}, requires
further analysis at $T=0$ as well as several metallic--, 
Kondo--type--extensions. Given the experimental facts about phase diagrams
in High$T_c$superconductors \cite{scalapino,LaSrCuO} the allowance for 
superconductivity in fermionic spin glass models is interesting too.\\
In more difficult models than those treated by us here, the new method of 
Fourier--Transformations in replica space \cite{temesvari} is hoped to 
facilitate further insight into the highly nontrivial 
$K=\infty$--solutions.\\
We finally note that an overlap distribution function for data clustering
shown in \cite{LootensBroeck} and interpreted as a pseudo $T=0$--problem in
a classical spin analogy, revealed, apart from the ratio discussed in
Eq.(\ref{10}), a remarkable similarity with the $(H=0)$--density of states.
It appears interesting to explore pseudo--$(T=0)$ neural network
problems \cite{gyrei,biehl,SFEdwards} as potential classical partners of 
fermionic spin glasses.\\

\section{Ward identity}
Conservation of charge and the corresponding continuity equation provide 
an exact relation which has a bearing on low energy two--particle 
excitations,
connecting them also with the one--particle excitations discussed before. 
Its feature, which is to connect (directly or indirectly) one--fermion 
propagators with vertices and higher order correlations, is particularly
important in the present context, since RPSB--breaking is thus transferred
into many--partice Green's functions. One may not be surprised to find 
RPSB in spin-- or charge--overlap correlations like 
$[\sigma^a(\tau)\sigma^b(\tau')]_{av}$ 
for example, but the less obvious strong effects derived before for 
replica--diagonal quantities will also be present. \\
The Ward identity for charge conservation will of course be most 
important for metallic models, since the replica--diagonal continuity 
equation 
\begin{equation}
div\{j^{aa}\}+\frac{\partial}{\partial 
t}(\sum_{\sigma}\overline{\psi}_{i,\sigma}^a(t)\psi_{i,\sigma}^a(t))=0
\end{equation}
where $div\{j^{aa}\}$ means the discretized divergence of the current 
operator which is given in terms of the fermion fields by 
\begin{equation}
j^{aa}(r_i)=-i e 
r_i\sum_{j,\sigma}\left[\overline{\psi}^a_{i,\sigma}t_{i,j}\psi^a_{j,\sigma}-
\overline{\psi}^a_{j,\sigma}t_{j,i}\psi^a_{i,\sigma}\right]
\end{equation}
links current-- and charge--correlations.\\
It does not seem necessary to invent replica--overlap currents and densities,
if not for the purpose of studying the existence of replica nondiagonal 
Green's functions ${\cal{G}}^{ab}$ involving nonhermitian realizations. 
It is clear that beyond replica permutation symmetry on the level of spins
there is the possibility to write, formally, another RPSB on the level of 
fermions. This is however not the subject of the present paper.\\
Let us recall the two--fermion propagators $K^{ab}(q,\omega)$ and the
density response function, which played a crucial role in the 
replica--symmetric theory of Anderson localization. Hopping disorder 
related diffusion of electrons was closely linked to the Ward identity.
Even if we would not question the stability of those theories against RPSB 
or perhaps other kinds of RSB, the theory of localization due to a random 
interaction will as well rely on the Ward identity. \\
The Ward identity for the insulating model can be
viewed as the $k=0$--part of the one for a metallic spin glass. Ignoring 
fermion momenta (anyway absent in the insulating case) one may employ
\begin{equation}
i\omega\hspace{.1cm}\Lambda_{\rho}(k=0,\epsilon+\omega,\epsilon)=
{\cal{G}}(\epsilon+\omega)-{\cal{G}}(\epsilon) 
\end{equation}
to get
\begin{equation}
\hspace{.2cm}lim_{\omega\rightarrow 0}\hspace{.1cm}\omega\hspace{.1cm}
lim_{k\rightarrow 0}\Lambda_{\rho}^{RA}(k,\epsilon+\omega,\epsilon)=
2\pi i\hspace{.1cm}\rho(\epsilon,\{q_r-q_{r-1}\})\nonumber
\end{equation}
shows that the Parisi form of the DoS, depending on all
$q_r-q_{r-1}$ or on $q(x)$ for $K=\infty$, also enters the vertex 
function. As usual the the three--point function involving the 
four--current vertex
$<T_{\tau}[a^{\dagger}_i(\tau)a_{i'}(\tau')j^{\mu}_{i''}(0)]>$
defines the two--legged quantity $\Lambda_{j^{\mu}}$.\\
The Ward identity for metallic or superconducting systems with allowance for
spin glass order and RPSB has the usual form
\begin{equation}
i \underline{k}\underline{\Lambda}_{j}(p+k,p;\epsilon+\omega,\epsilon)+
i\omega\Lambda_{\rho}(p+k,p;\epsilon+\omega,\epsilon)=
{\cal{G}}(p+k,\epsilon+\omega)-{\cal{G}}(p,\epsilon),
\end{equation}
where ${\cal{G}}$ denotes again the disorder averaged and 
replica--diagonal Green's function, which incorporates even at $T=0$ 
strong RPSB--effects.\\
It is very tempting to mention the consequences of a representation 
$(\ref{72})$, although we do not yet know, whether it holds beyond the 
insulating model. It appears likely that it may hold within the Q--static
approximation (since the Parisi type average given here involves
quantities that are static in the insulating case but become 
time--dependent in the metallic case). 
Thus, within the limits of its validity, this representation could be 
generalized in a rather simple way to the 
metallic or superconducting cases at zero temperature, and one would obtain 
the $(T=0)$ Ward identity in the form similar to the one of clean systems 
but including an disorder average with constrained (cutoff) Parisi type 
distributions (depending on te number of RSB--steps). Of course all order 
parameters would have to be determined selfconsistently for each 
particular case.
Inferring the form of the $K=1$ propagator, Eq.(\ref{72}), the vertex (with
external legs)
$\Lambda$ depends on the Parisi order parameters as (in shorthand notation)
\begin{equation}
i q_{\mu}\Lambda_{\mu}(p+k,p;\epsilon+\omega,\epsilon)=\int_2^G\frac{\int_1^G
\Theta(\tilde{H}_0)e^{a\tilde{H}_0}\left[\sum_{s=\pm1}
\{\frac{1}{i\epsilon+i\omega-\epsilon_{p+k}+s(\tilde{H}_0+\overline{\chi})}
-\frac{1}{i\epsilon-\epsilon_p+s(\tilde{H}_0+\overline{\chi})}\}
\right]}{\int_1^G e^{a\tilde{H}_0}}
\end{equation}
This will also occur in metallic spin glasses, whence diffusive modes and
conductivity are expected to depend on Parisi symmetry breaking.

\section{Mesoscopic Fluctuations in Quantum Spin Glass Phases}
Spin glasses have been discussed in the context of mesoscopic systems 
\cite{mydosh}. 
Nanostructuring of materials with frustrated magnetic 
interactions promise a large field of applications \cite{mydosh}.
We contributed a recent example of a semimagnetic layered semiconductor 
system \cite{chud}.\\
Mesoscopic fluctuations have been studied intensively for electronic systems
with random potentials. The possibility of RPSB was ignored or assumed to 
be irrelevant or nonexisting. We provide here the technique that helps to
reconsider this question. This section however briefly evokes only a few
examples of mesoscopic fluctuations in spin glass systems.\\
A typical example is given by statistical fluctuations of the density
of states. Multi--valley correlations described by 
\begin{equation}
C_{\rho}^{(k)}\equiv 
\overline{\rho_{a_1}(E_1)\rho_{a_2}(E_2)...\rho_{a_k}(E_k)},
\end{equation}
where the $a_i$ denote distinct replicas, do not factorize within the
spin glass phase. For the choice $|E_k|=E<E_g(H)$ we obtain
\begin{equation}
C_{\rho}^{(k)}(E)=
(\rho(E))^k\frac{B(\frac{k}{2} d_-^E,\frac{k}{2} d_+^E)}
{(B(\frac{1}{2}d_-^E,\frac{1}{2}d_+^E))^k}
\end{equation}
with $d_{\pm}^E\equiv \frac{E_g(H)\pm E}{E_g(H)}$ and Beta--function B.
These multi--valley correlations of arbitrary order k also vanish 
exponentially everywhere in the gap as $T\rightarrow 0$ and,
much weaker than the averaged DoS though,
they increase as either gap edge is approached at fixed low temperatures.\\
%
%

\section{Appendix A: Quantum field theory and decoupling procedure}
\subsection{A1: the fermionic Ising spin glass}
The gaussian average over random fluctuations in the magnetic coupling 
$J_{ij}$ renders the thermodynamic potential 
$\Omega=Tr e^{-\beta(H-\mu N)}$ of the fermionic Ising spin glass as
\begin{equation}
\Omega=\int 
exp[\sum_{ij}\int_{\tau,\tau'}M_J(i-j)\left[\sum_a 
X^{aa}_i(\tau,\tau')X^{aa}_j(\tau',\tau)
+2\sum_{a<b}X^{a,b}_i(\tau,\tau')X^{ba}_j(\tau',\tau)\right]], 
\end{equation}
where
\begin{equation}
X^{ab}_i(\tau,\tau')\equiv\sum_{\lambda=\pm 1} 
\overline{\psi}^a_{i\lambda}(\tau)
\lambda\psi^a_{i\lambda}(\tau)\sum_{\lambda'=\pm 1}
\overline{\psi}^b_{i\lambda'}(\tau')\lambda'\psi^b_{i\lambda'}(\tau')
\end{equation}
We decouple this eight--fermion correlation by a matrix field 
$Q_i^{a,a'}(\tau,\tau')$. 
Since the fermionic Ising model does neither have
spin-- nor charge--quantum dynamics, and since $\underline{\underline{Q}}$ 
represents a coarse--grained
spin overlap field with the same average as the X--field, its static and 
spatially homogeneous saddle point solution, which has to be determined 
selfconsistently, is extracted by
\begin{equation}
Q_i^{a,a'}(\tau,\tau')\equiv\Lambda^{a,a'} + \delta Q_i^{a,a'}(\tau,\tau')
\end{equation}
where the saddle point matrix $Q_{sp}\equiv\Lambda$ turns out to be of the 
usual Parisi form plus a diagonal matrix taking care of 
$(\sigma^a)^2$--averages, hence
\begin{equation}
\Lambda^{a,a'}=q_{_{_{Parisi}}}+\tilde{q}\hspace{.1cm}\underline{\underline{1}}
\end{equation}
The infinite--range fermionic Ising spin glass is then described by
\begin{equation}
\Omega=lim_{n\rightarrow 0}\frac{1}{n}\left[<Z_n>-1\right]
\end{equation}
with
\begin{equation}
<Z_n>=\int{\cal{D}}Q\int{\cal{D}}m e^{-nN {\cal{A}}(Q,m)}
\end{equation}
The saddle point contribution for the infinite system ($N\rightarrow\infty$)
can be extracted by
\begin{equation}
=\int {\cal{D}}\Psi 
exp\left[
-\frac{N}{4}(\beta J)^2\sum_a(q^{aa})^2+
\frac{1}{2}q^{aa}\int_{\tau}\int_{\tau'}X_{\tau\tau'}^{aa}       
\right]
exp(\left[-\frac{N}{2}(\beta J)^2\sum_{a<b}(q^{ab})^2+
J^2\sum_{a<b}q^{ab}\int_{\tau}\int_{\tau'}X_{\tau\tau'}^{ab} 
\right], 
\end{equation}
while the fluctuation part is expanded in terms of the fluctuation fields
$\delta Q$. Time--independence of the spin correlations for 
the Ising case including charge interactions and grand canonical description, 
allows to perform time integrations on the fluctuation fields. Hence only the
$\delta Q(\omega=0,\omega=0)$ components remain. This is different in the
metallic case and the separation of a dynamic saddle point together with
the dynamic fluctuation theory in terms of fields $\delta Q(\omega,\omega')$
was described in \cite{sro}.\\
The second class of decouplings, needed to reduce 4th--order Grassmann field
products to the integrable 2nd order products, must take into account the 
RPSB in the Q--matrix. With each step K of RPSB te number of new decoupling 
fields increases. The procedure is clearly seen for example at $K=2$, where
\begin{eqnarray}
<Z_n>&=&e^{-\frac{N}{4}\beta^2 J^2 Tr Q^2_{_{Parisi}}}\prod\int 
d\overline{\psi}d\psi 
exp\left[(\overline{\psi}i\epsilon\psi)+\frac{1}{2}\beta^2 J^2 
Tr(\underline{\underline{Q}}_{_{Parisi}}
\underline{\underline{X}})\right]\\
&=&\prod_{\alpha_3}\int^G_{z_3^{(\alpha_3)}}...\prod_{\alpha_0}
\int^G_{z_0^{(\alpha_0)}}\prod\int d\overline{\psi}d\psi 
exp\{
\sum^{\alpha_3\frac{m_3}{m_2}}_{\alpha_2=(\alpha_3-1)\frac{m_3}{m_2}+1}
\sum^{\alpha_2\frac{m_2}{m_1}}_{\alpha_1=(\alpha_2-1)\frac{m_2}{m_1}+1}   
\sum^{\alpha_1\frac{m_1}{m_0}}_{\alpha_0=(\alpha_1-1)\frac{m_1}{m_0}+1}
\sum_{i,\sigma,\epsilon_l}\nonumber\\
& &\hspace{-1.5cm}\overline{\psi}^{(\alpha_0),l}_{i,\sigma}
\left[i\epsilon_{_l}+\sigma(h_i^{(\alpha_0)}+J\sqrt{q_0-q_1}z_0^{(\alpha_0)}
+J\sqrt{q_1-q_2}z_1^{(\alpha_1)}+J\sqrt{q_2-q_3}z_2^{(\alpha_2)}+
J\sqrt{q_3-q_4}z_3^{(\alpha_3)})\right]
\psi_{i\sigma}^{(\alpha_0),l}\}, 
\label{82}\nonumber
\end{eqnarray}
where the decoupling fields carry a Parisi block index 
$k$ (summations over $\alpha_k$ run over replicas pertaining to this block) 
and $q_{_0}\equiv\tilde{q}$, $z_{_0}\equiv y^a$. 
The generalization to arbitrary K is now obvious. Generating 
fermion fields $\eta,\overline{\eta}$ can be added to allow for the 
calculation of Green's functions. Let us define 
\begin{eqnarray}
\prod_{\mu}(\prod_{\alpha_{\mu}}\int_{_{z_{\mu}^{(\alpha_{\mu})}}})
&\equiv&\int^G_{_{z_{k+1}^{(\alpha_{k+1})}}}
\prod_{\alpha_k}\int^G_{_{z_k^{(\alpha_k)}}}...
\prod_{\alpha_0}\int^G_{_{z_0^{(\alpha_0)}}}\\
\prod_{\mu}\left[\sum_{\alpha_{\mu}}\right]
&\equiv&\sum_{\alpha_k=(\alpha_{k+1}-1)\frac{m_{k+1}}{m_k}+1}^{\alpha_{k+1}
\frac{m_{k+1}}{m_k}}\sum_{\alpha_{k-1}=
(\alpha_{k}-1)\frac{m_k}{m_{k-1}}+1}^{\alpha_k
\frac{m_k}{m_{k-1}}}
.....
\sum_{\alpha_0=(\alpha_1-1)\frac{m_1}{m_0}+1}^{\alpha_1
\frac{m_1}{m_0}}.
\end{eqnarray}
The generating functional for an arbitrary number K of RPSB--steps is 
then obtained as 
\begin{eqnarray}
\Xi_n(\eta,\bar{\eta})&=&e^{-\frac{N}{4}\beta^2 J^2 Tr
Q_{Parisi}^2}\prod[\prod\int_{z_{\gamma}^{(\alpha_{\gamma})}}]
\prod\int d\bar{\psi}d\psi\\
& &Exp\left[\prod_{\gamma}^{K}\left[\sum_{\alpha_{\gamma}}\right]
\sum_{i,\sigma,\epsilon_l}\bar{\psi}^{\alpha_0 l}_{i,\sigma}
g_{\sigma}^{-1}(\epsilon_l,\{z_{\gamma}^{\alpha_{\gamma}}\})
\psi^{\alpha_0 l}_{i,\sigma}+
\eta_{i,\sigma}^{\alpha_0 l}\bar{\psi}_{i,\sigma}^{\alpha_0 l}+
\bar{\eta}_{i,\sigma}^{\alpha_0 l}
\psi_{i,\sigma}^{\alpha_{0}l}\right]\nonumber
\end{eqnarray}
where the statistically fluctuating Green's function g is given by
\begin{equation}
g_{\sigma}^{-1}(\epsilon_l,\{z_{\gamma}^{\alpha_{\gamma}}\})=
[g_0^{-1}(\epsilon_l)+
\sigma\tilde{H}(\{z_{\gamma}^{(\alpha_{\gamma})}\})]^{-1}.
\end{equation}
The bare propagator $g(\epsilon_l)=1/(i\epsilon_l+\mu)$
becomes spatially dependent in metallic models, while all of the Ising spin 
glass complications including Parisi RPSB are taken care of by the fields 
$z_{\gamma}$. A shift of the fields leads to the result
\begin{eqnarray}
Exp(\Phi(\eta,\overline{\eta}))&=&(\prod\int d\psi\overline{\psi})Exp(\sum
\overline{\psi}^{al}_{\sigma}g^{-1}_{\sigma}(\epsilon_l,
\{z_{\gamma}\})\psi^{al}_{\sigma}+\eta\overline{\psi}-\overline{\eta}\psi)\\
&=&Exp(\Phi(0,0))Exp(-\sum\overline{\eta}^{al}_{\sigma}
g_{\sigma}(\epsilon_l,\{z_{\gamma}\})\eta^{al}_{\sigma}),
\end{eqnarray}
which shows that the averaged Green's function is obtained by the given
K+1 gaussian integrations over the fluctuation fields, contained in the 
function g, according to
\begin{equation}
{\cal{G}}_{\sigma}(\epsilon_l)=\frac{\int^G_{z_K}(\prod\int^G_{z_{K-1}})...
(\prod\int^G_{z_1})(\prod\int^G_{z_0})
Exp\left[\Phi(0,0)\right]g_{\sigma}(\epsilon_l,\{z_{\gamma}\})}{\int^G_{z_K}
(\prod\int^G_{z_{K-1}})...(\prod\int^G_{z_0})Exp\left[\Phi(0,0)\right]}.
\label{greensfunction}
\end{equation}
Apart from the gaussian weight indicated by the upper index G, one needs
the regularized (provided the continuous time formalism is used; an example
for the application of the discrete time formalism is given in Appendix 2 
for the local Hubbard limit) solution of
\begin{eqnarray}
Exp\left[\Phi(0,0)\right]&=&
\prod_{\{b_{\alpha}\}}\prod_{\epsilon_l}
\left[\epsilon_l^2+\mu^2+\tilde{H}^2(z_0,z_1,...,z_K)
\right]|_{\it reg}\\
&=&
\prod_{\{b_{\alpha}\}}\left[2 
cosh(\beta\tilde{H}(z_0,z_1,...,z_K))+2 cosh(\beta\mu)\right]
\end{eqnarray}

\subsection{Appendix 2: The local Hubbard limit with random U}
The mapping between the Green's function of the fermionic Ising spin glass 
and the one for a Hubbard interaction with random positive U and 
properly  chosen mean value of U, both taken at $T=0$ and at 
half--filling is surprising in several respects. A search for more general 
relations is necessary. A general method is of course the comparison of the 
field theories. A closer look on the field theoretic representation of a random 
U Hubbard interaction with cutoff gaussian $\delta U$ distribution is 
needed.\\
The decoupling of the Hubbard interaction, using the operator identity 
$n_{\uparrow}n_{\downarrow}=\frac{1}{4}((\sum_{\lambda}n_{\lambda})^2-
(\sigma^z)^2)$, was given by
Vollhardt \cite{vollhardt}. Let us first reconsider the Grassmann integrals
using the discrete times (introduced by time slicing), since the Hubbard 
interaction is one of the nice examples that allows to carry through a 
regularisation--free formulation.\\
The partition function reads
\begin{eqnarray}
Z&=&lim_{M\rightarrow\infty}\prod_{\sigma}\prod_{k=0}^{M-1}\int 
d\overline{\psi}_{\sigma,k}d\psi_{\sigma,k}exp\left[\epsilon 
U\sum_{k=0}^{M-2}
\overline{\psi}_{\uparrow,k+1}\overline{\psi}_{\downarrow,k+1}
\psi_{\uparrow k}\psi_{\downarrow k}+\epsilon U\overline{\psi}_{\uparrow 0}
\overline{\psi}_{\downarrow 0}\psi_{\uparrow M-1}\psi_{\downarrow 
M-1}\right]\\ 
& &\hspace{-1cm}exp\left[\epsilon\mu\sum_{k=0}^{M-2}\sum_{\sigma}
\overline{\psi}_{\sigma k+1}\psi_{\sigma k}
-\epsilon\mu\sum_{\sigma}\overline{\psi}_{\sigma 0}
\psi_{\sigma M-1}-\sum_{k=0}^{M-2}\sum_{\sigma}\overline{\psi}_{\sigma k+1}
(\psi_{\sigma k+1}-\psi_{\sigma k})-\sum_{\sigma}
\overline{\psi}_{\sigma 0}(\psi_{\sigma 0}-\psi_{\sigma M-1})\right]\nonumber
\end{eqnarray}
where infinitesimal time steps $\epsilon\equiv \frac{\beta}{M}$ are employed.
Using the Grassmann equivalent of the operator identity given above, the
interaction term is decoupled at any time instant by
\begin{equation}
e^{\epsilon U\overline{\psi}_{\uparrow k+1}\overline{\psi}_{\downarrow
k+1}\psi_{\uparrow k}\psi_{\downarrow k}}= 
\int^{\infty}_{-\infty}\frac{d\alpha_k}{\sqrt{2\pi}}e^{-\frac{1}{2}\alpha_k^2}
e^{i\sqrt{\frac{\epsilon U}{2}}\alpha_k\sum_{\sigma}\overline{\psi}_{\sigma 
k+1}\psi_{\sigma k}}+\int^{\infty}_{-\infty}\frac{d\gamma_k}{\sqrt{2\pi}}
e^{-\frac{1}{2}\gamma_k^2}e^{\sqrt{\frac{\epsilon U}{2}}\gamma_k\sum_{\sigma}
\overline{\psi}_{\sigma k+1}\sigma\psi_{\sigma k}}
\end{equation}
As for continuous times \cite{vollhardt} all integrals can be performed 
exactly. The bilinear exponent is expressed in terms of a matrix 
$(B^{\sigma,\sigma'}_{k,k'})$ whence 
\begin{equation}
Z=lim_{M\rightarrow\infty}\prod\int{\cal{D}}\psi\int_{\alpha_k}^G
\int^G_{\gamma_k}e^{-\overline{\Psi}B\Psi}=
lim_{M\rightarrow\infty}\prod\int^G_{\alpha_k}\int^G_{\gamma_k}det B
\end{equation}
The determinant is found as
\begin{equation}
det B(U)=1+\left[\sum_{\lambda}\prod_k+\prod_k\prod_{\lambda}\right]
\left[1+\epsilon\mu+\sqrt{\frac{\epsilon U}{2}}
(i\alpha_k+\lambda\gamma_k)\right]
\end{equation}
The remaining integrations over the charge-- and spin--decoupling fields
are readily evaluated to reproduce the result $Z=1+2 e^{\beta\mu}
+e^{\beta(2\mu-U)}$ for each site.\\
This result being trivially derivable within the operator formalism, the 
filed theory obviously complicates the derivation. Nevertheless field 
theories provide the best way to find general relations, for which we are 
looking here. In fact, since we know that the infinite--range fermionic spin 
glass cannot be represented at all temperatures by one zero--dimensional
random U Hubbard model, we are looking for the differences in those models
field theoretic descriptions. Before discussing the effect of cutoff 
gaussian distributions in replicated extensions of the given partition 
function, let us the generating functional in the discrete time formalism.\\
The generating functional
\begin{equation}
\Xi(\eta,\overline{\eta})=ln\left[lim_{_{M\rightarrow\infty}}
\prod_{\sigma,k}\int
{\cal{D}}\Psi\int^G_{\alpha_k}\int^G_{\gamma_k}e^{-\overline{\Psi}B\Psi
+\eta\overline{\psi}-\overline{\eta}\psi}\right]
=ln\left[lim_{_{M\rightarrow\infty}}\prod_{\sigma,k}det B 
e^{\overline{\eta}B^{-1}\eta}\right] 
\end{equation}
only requires to invert matrix B. This yields
\begin{equation}
{\cal{G}}(\tau_f-\tau_i)=-\frac{1}{Z}\prod_k\int_{\alpha_k}^G
\int_{\gamma_k}^G\prod_{l=k_i}^{k_f}B^{(\sigma)}_{l+1,l}det B^{-\sigma},
\end{equation}
This is readily evaluated to be (for $\tau_f>\tau_i$)
\begin{eqnarray}
{\cal{G}}_{\uparrow}(\tau_f-\tau_i)=-\frac{1}{Z}lim_{M\rightarrow\infty}
\prod_{k=0}^{M-1}\int_{\alpha_k}^G\int_{\gamma_k}^G\prod_{k_i}^{k_f-1}
\left[1+\epsilon\mu+\sqrt{\frac{\epsilon U}{2}}(i\alpha_k+\gamma_k)\right]
\left[1+\prod_0^{M-1}(1+\epsilon\mu+\sqrt{\frac{\epsilon 
U}{2}}(i\alpha_k-\gamma_k))\right]\nonumber\\ 
& &\hspace{-17.5cm}
=-\left[e^{\mu(\tau_f-\tau_i)}+e^{\beta\mu}e^{(\mu-U)(\tau_f-\tau_i)}\right]/Z
\hspace{9.3cm}(133)\nonumber
\end{eqnarray}
Since we have now at hand the field theory which correctly reproduces the
known results for Z and Green's function, we know that the replica formalism
applied to a random U local Hubbard limit with the incomplete gaussian 
distribution obtained in (\ref{50}) must reproduce the average 
of ln Z and G. 
It will be interesting and necessary to study in more detail the field theory
of the random Hubbard model with an incomplete gaussian 
distribution and its relation with the metallic spin glass.

\end{document}